\documentclass[11pt,twocolumn]{article}
\usepackage{url}
\usepackage[colorlinks = true, linkcolor = blue, urlcolor = blue, citecolor = blue, anchorcolor = blue]{hyperref}
\usepackage{amsmath}
\usepackage{amsfonts}
\usepackage{amsthm}
\usepackage{graphicx}
\usepackage{epstopdf}
\usepackage{subfig}
\usepackage[font=scriptsize]{caption}
\usepackage{cite}
\usepackage{color}
\usepackage{tikz}
\usepackage[ruled,vlined]{algorithm2e}
\newcommand{\ud}{\,\mathrm{d}}
\usepackage[paper=a4paper,left=15mm,right=14mm,top=10mm,bottom=15mm]{geometry}
\title{Breaking the Limits -- Redefining the Instantaneous Frequency}
\author{Pushpendra Singh$^{1,2,}$\footnote{Author's E-mail address: \texttt{spushp@gmail.com} (P. Singh); \texttt{pushpendrasingh@iitkalumni.org}}
\\{\normalsize $^{1}$Department of Electrical Engineering, Indian Institute of Technology Delhi, India}\\
{\normalsize $^{2}$Jaypee Institute of Information Technology - NOIDA, India}\\}
\providecommand{\keywords}[1]{\textbf{\textit{Keywords:}} #1}
\date{}
\begin{document}
\maketitle
\begin{abstract} % prsa
The Carson and Fry (1937) introduced the concept of variable frequency as a generalization of the constant frequency. The  instantaneous frequency (IF) is the time derivative of the instantaneous phase and it is well-defined only when this derivative is positive. If this derivative is negative, the IF creates problem because it does not provide any physical significance. This study proposes a mathematical solution and eliminate this problem by redefining the IF such that it is valid for all monocomponent and multicomponent signals which can be nonlinear and nonstationary in nature. This is achieved by using the property of the multivalued inverse tangent function. The efforts and understanding of all the methods based on the IF would improve significantly by using this proposed definition of the IF. We also demonstrate that the decomposition of a signal, using zero-phase filtering based on the well established Fourier and filter theory, into a set of desired frequency bands with proposed IF produces accurate time-frequency-energy (TFE) distribution that reveals true nature of signal. Simulation results demonstrate the efficacy of the proposed IF that makes zero-phase filter based decomposition most powerful, for the TFE analysis of a signal, as compared to other existing methods in the literature.
\end{abstract}

\section{INTRODUCTION}
The time-domain representation and the frequency-domain representation are two classical representations of a signal. In
both domains, the time ($t$) and frequency ($f$) variables are mutually exclusive~\cite{blBB}. The Time-Frequency Distributions (TFDs) representation on the other hand, provides localized signal information in time and frequency domain. The TFD provides insight into the complex structure of a signal consisting of
several components. There exist many types of time-frequency analysis methods such as short-time Fourier transform (STFT), Gabor transform, wavelet transforms, Wigner-Ville distribution.

The Carson and Fry~\cite{th19} introduced the concept of variable frequency, required to the theory of frequency modulation (FM), as a generalization of the definition of constant frequency. Moreover, the nonstationary nature of the signals and nonlinear systems require the idea of instantaneous frequency (IF). The IF is the basis of the TFD or time-frequency-energy (TFE) representation and analysis of a signal. The IF is a practically important parameter of a signal which can reveal the underlying process and provides explanations for physical phenomenon in many applications such as vibration, acoustic, speech signal analysis~\cite{rslc9}, meteorological and atmospheric applications~\cite{rslc1}, seismic~\cite{rslc9}, radar, sonar, solar physics, structural engineering, communications, health monitoring, biomedical and medical applications~\cite{th46}, cosmological gravity wave and financial market data analysis.

The IF is the time derivative of the instantaneous phase (IP) and it is well-defined only when this derivative is positive. If this derivative is negative, it creates problem because it does not provide any physical significance. In order to avoid this problem, recently many nonlinear and nonstationary signal representation, decomposition and analysis methods, e.g. empirical mode decomposition (EMD) algorithms~\cite{rslc1,rslc2,rslc3,rslc4,rslc5,rslc6,rslc7,co1}, synchrosqueezed wavelet transforms (SSWT)~\cite{rslc71}, variational mode decomposition (VMD)~\cite{rslc72}, eigenvalue decomposition (EVD)~\cite{rslc73}, empirical wavelet transform (EWT)~\cite{rslc74}, sparse time-frequency representation~\cite{co2}, time-varying vibration decomposition~\cite{co3}, resonance-based signal decomposition~\cite{co4} and Fourier decomposition methods (FDM)~\cite{rslc8,rslc9,rslc101,rslc11,rslc12,rslc13} based on the Fourier theory, are proposed. The Fourier theory is the only tool for spectrum analysis of a signal and the FDM has established that it is a superior tool for nonlinear and nonstationary time series analysis. The main objective of all these methods is to obtain the signal representation such that the IF of a signal understudy is always positive.

Unlike these decomposition methods, the IF proposed in this study does not necessitate to decompose a signal into a set of narrow band components, which satisfy certain properties, to generate the TFE distribution of a signal. That is, without any decomposition the TFE distribution of a signal can be obtained. It also provide freedom and potential to decompose a signal into a set of desired (preferably orthogonal or LINOEP~\cite{rslc4}) frequency bands by a zero-phase filtering approach to obtain TFE distribution of a signal. All these features are obtained by redefining the IF when it is negative and thus defining the IF for both monocomponent and multicomponent signals. In order to redefine the IF, we use the fact that inverse tangent is a \emph{multivalued} (i.e. one-to-many mapping) function. So defined IF of a signal is always positive and valid for any signal.

The main contributions as well as some important observations of this study are as follows:
\begin{enumerate}
\itemsep0em
\item We use the conventional definition of the IF when it is positive, if it is negative then redefine the IF to make it positive. Thus, the proposed IF is always positive and valid for all monocomponent as well as multicomponent signals, which can be nonstationary and nonlinear in nature.
\item Since many decades, there has been a general understanding in the literature, e.g.~\cite{rslc1,blBB,rslc2,rslc3,rslc73,th4,th411,rs22}, that the Fourier theory (due to linearity, periodicity and stationarity) is not suitable for nonstationary signal analysis. This proposed IF provides an elegant way to use the Fourier and filter theory based zero-phase filtering for the decomposition of a signal into a set of suitable bands with desired cutoff frequencies (e.g. divide complete bandwidth of a signal into a set of sub-bands of equal bandwidth or dyadic sub-bands). In order to validate this, in the study, we have used only the Fourier and finite impulse response (FIR) filter theory based decomposition (except when comparing with EMD, FDM, wavelet transform and conventional non zero-phase FIR filtering) to obtain TFE analysis of a signal.
\item We demonstrate that the zero-phase discrete Fourier transform (DFT) filter-bank based decomposition of a signal produces orthogonal components, however it is more natural to obtain LINOEP components with zero-phase FIR filter based decomposition. The both set of orthogonal and LINOEP vectors preserve the energy in decomposition, and present similar TFE distribution of a signal.
\item The proposed method, using the Hilbert spectrum, produces average frequencies in the TFE distribution with good time resolution when the envelope of signal is smooth. However, if envelope of a signal is fluctuating randomly and rapidly, e.g. the Gaussian white noise and Earthquake time series, the TFE plot presents good time and frequency resolution.
\item We demonstrate that the different decomposition methods are producing the different TFE distributions of a signal. Using the proposed method, when a signal is decomposed into more numbers of orthogonal or LINOEP narrow bands, true frequencies present in the signal are revealed, frequency resolution also increases while the time resolution reduces marginally.
\end{enumerate}
Thus, this study presents a new paradigm for nonlinear and nonstationary data analysis. Moreover, this work independently and uniquely resolves some misconceptions that have grown with regard to the significance of the Fourier theory and its usefulness in the representation and analysis of nonstationary signal, and also supports the results of other related studies~\cite{rslc9,rslc101}.
This paper is organized as follows: The proposed methodology is presented in Section~\ref{meth}. Simulation results and discussion are presented in Section~\ref{simre}. Section~\ref{con} presents conclusion of the work.
\section{METHODS}\label{meth}
The concept of variable frequency with application to the theory of frequency modulation (FM) is proposed in~\cite{th19}, and it is postulated that the notion of IF is a generalization of the definition of constant frequency.
A definition of the IF by analyzing an expression for simple harmonic motion (SHM) is considered in~\cite{th20} as
\begin{equation}
x_{\text{\tiny SHM}}(t)=a \cos\Big[\int_{0}^{t} 2\pi f(t) \ud t+ \theta\Big]=a \cos(\psi(t)), \label{Ch1_eq5}
\end{equation}
where the argument of the cosine function is the phase $\psi(t)=[\int_{0}^{t} 2\pi f(t) \ud t+ \theta]$. This leads to the definition of instantaneous frequency~\cite{th20}
\begin{equation}
f(t)=\frac{1}{2\pi} \frac{\ud\psi(t)}{\ud t}. \label{Ch1_eq6}
\end{equation}
The concept of instantaneous frequency was enhanced in~\cite{th21} where a method for generating a unique complex signal $z(t)$ from a real signal $x(t)$ and its Hilbert transform $\hat{x}(t)$ was proposed. This method obtains
\begin{equation}
        z(t) =x(t)+j\hat{x}(t) = a(t)e^{j\phi(t)}, \label{Ch1_eq7}
\end{equation}
where
\begin{equation}
 \left.\begin{aligned}
        a(t) & =[x^2(t)+\hat{x}^2(t)]^{1/2}\ge 0, \\
         \phi(t) & =\tan^{-1}[\hat{x}(t)/x(t)]
       \end{aligned}
 \right\} \label{Ch1_eq71}
\end{equation}
and $z(t)$ is the Gabor's complex signal (well-known as the analytic signal) and $\hat{x}(t)$ is the Hilbert Transform (HT) of $x(t)$, defined as
\begin{equation}
\hat{x}(t)= \text{p.v.} \int_{-\infty}^{\infty} \frac{x(\tau)}{\pi (t-\tau)} \ud \tau = \text{p.v.} \left[x(t)*\frac{1}{\pi (t)}\right], \label{Ch1_eq8}
\end{equation}
where p.v. denotes the Cauchy principal value of the integral~\cite{th22} and $*$ denotes convolution operation.
The work done in~\cite{th19} and~\cite{th21} was unified in~\cite{th23} to define the IF of a signal $x(t) = a(t) \cos(\phi(t))$ as
\begin{equation}
\omega(t)= 2\pi f(t) =\frac{\ud}{\ud t} \left(\text{arg} [z(t)]\right), \label{Ch1_eq9}
\end{equation}
where $z (t)$ is the analytic signal given by~\eqref{Ch1_eq7} and $\text{arg} [z(t)]=\phi(t)$ as defined in \eqref{Ch1_eq71}. The IF, as defined in \eqref{Ch1_eq9}, provides physical meaning only when it is positive and it becomes meaningless when it is negative~\cite{rslc1,rslc2,rslc3,rslc4,rslc6,rslc71,rslc72,rslc73,rslc8,rslc9,th4,th411}. This is where we improve and provide the solution to make the IF positive for all time $t$. From~\eqref{Ch1_eq71} and~\eqref{Ch1_eq9}, one can obtain the TFE distribution by 3-D plot of $\{t, f(t), a^2(t)\}$.
However, it is to be noted that the frequencies, of signal $x(t)=a(t)\cos(\phi(t))$, obtained by the Fourier transform and Hilbert spectrum~\eqref{Ch1_eq9} are same only if $a(t)$ is a constant, otherwise Hilbert spectrum gives frequencies of signal $\cos(\phi(t))$ and not of signal $x(t)$.

Before evaluating the time derivative of phase in~\eqref{Ch1_eq9}, phase unwrapping is necessary to ensure that all appropriate multiples of $2\pi$ have been included in phase $\phi(t)$. Phase unwrap operation corrects the radian phase angles by adding multiples of $\pm 2\pi$ when absolute jumps between consecutive elements of a phase vector are greater than or equal to the default jump tolerance of $\pi$ radians~\cite{matlabweb}. This is being done for phase delay and IF determination by all the methods available in literature.

As is well-known that the tangent is a surjective (many-to-one mapping) function. The domain, range and period of $\tan(x)$ are $\{x|x\ne \frac{\pi}{2}+n\pi, \forall n \in \mathbb{Z}\}$, all real numbers $\mathbb{R}=(-\infty,\infty)$ and $\pi$, respectively. %i.e., $\tan(\pi+x)=\tan(x)$.
The inverse tangent is the \emph{multivalued} function. The domain of $\tan^{-1}(x)$ is all real numbers, $\mathbb{R}$, and range is $(-\frac{\pi}{2},\frac{\pi}{2})$. If $z=x+jy$, then the range of $\tan^{-1}(y/x)$ is $(-{\pi},{\pi}]$, sign of $x$ and $y$ is used to determine the specific quadrant.

In order to obtain the IF positive for all the time, we consider the discrete-time signal processing which is the only practical way to process data by a processor. The discrete time version of the equations~\eqref{Ch1_eq7},~\eqref{Ch1_eq71},~\eqref{Ch1_eq8} and~\eqref{Ch1_eq9} are~\cite{th411}
\begin{subequations}
\begin{align}
        z[n] & =x[n]+j\hat{x}[n] = a[n]e^{j\phi[n]}, \label{sub1}\\
        a[n] & =\left[x^2[n]+\hat{x}^2[n]\right]^{1/2}\ge 0,\label{sub2} \\
        \phi[n] & =\tan^{-1}\left(\hat{x}[n]/x[n]\right),\label{sub3}\\
        \hat{x}[n] & =x[n]*\left(\frac{1-\cos(\pi n)}{\pi n}\right),\label{sub4}\\
    \text{ and } \omega[n] & =\phi_d[n], \label{sub5}
\end{align}\label{Ch1_eq11}
\end{subequations}
\noindent respectively, where the differentiation in discrete-time can be approximated by~\cite{th411} forward finite difference (FFD), $\phi_d[n]=\big(\phi[n+1] -\phi[n] \big)$, or backward finite difference (BFD), $\phi_d[n]=\big(\phi[n] -\phi[n-1] \big)$ or central finite difference (CFD), $\phi_d[n]=\big(\phi[n+1] -\phi[n-1] \big)/2$. It is to be noted that the phase in~\eqref{sub3} is computed by the function, $\textup{atan2}(\hat{x}[n],x[n])$, which produces the result in the range $(-\pi, \pi]$ and also avoids the problems of division by zero.

By considering the phase unwrapping fact and \emph{multivalued} nature of the inverse tangent function, we redefine the IF $\omega[n]$, defined in~\eqref{sub5} as a discrete-time version of~\eqref{Ch1_eq9}, as
\begin{equation}
 \omega[n] =
  \begin{cases}
    \phi_d[n],  \text{ if } \phi_d[n] \geq 0, \\
   \phi_d[n]+\pi,  \text{ otherwise,}
  \end{cases}\label{Ch1_eq12}
\end{equation}
which makes the IF positive for all time $n$. This small and trivial but extremely important fact has been illusive for many decades.
This IF would not only significantly improve the computational efforts (e.g. sifting process in EMD algorithms) and understanding of any method which uses the IF, but also provides an elegant solution in mathematical terms to use the Fourier and filter theory based zero-phase filtering for nonstationary signal decomposition and TFE analysis. The mathematical validity of this solution can easily be seen by the fact that $\phi[n] =\tan^{-1}\left(\hat{x}[n]/x[n]\right)$ and $\phi[n]+kn\pi=\tan^{-1}\left(\hat{x}[n]/x[n]\right), \forall k, n \in \mathbb{Z}$ (due to periodicity of tangent function, i.e. $\tan(\phi[n])=\tan(\phi[n]+kn\pi) $). In solution \eqref{Ch1_eq12}, we have taken $k=0$ if $\phi_d[n]\ge 0$, otherwise $k=1$, thus we can write
\begin{equation}
 \omega[n]=\phi_d[n]+k\pi. \label{Ch1_eq12_1}
\end{equation}
The  equation~\eqref{Ch1_eq12_1} is also valid when $\phi[n]\in (-\pi,\pi]$ is a wrapped phase and in this case we select the value of $k\in \mathbb{Z}$ such that $\omega[n]\in [0,\pi]$. Thus, the proposed IF $\omega[n]$ in~\eqref{Ch1_eq12} or~\eqref{Ch1_eq12_1} is estimated by using the multivalued property of the inverse tangent function which provides basis to ensure that the instantaneous phase function $\phi[n]$ defined in~\eqref{sub3} is an \emph{increasing} (or a \emph{nondecreasing}) function, i.e. $\phi[n+1]\ge \phi[n], \forall n$.

A dual to IF~\eqref{Ch1_eq12}, the group delay (GD) is defined as the negative frequency derivative of the phase in the Fourier domain. % i.e. $\tau_{gd}=-\frac{\ud}{\ud f} \phi(f)$.
It measures the relative delay of different frequencies from the input to the output in a system. Thus, similar to \eqref{Ch1_eq12}, we can modify the GD definition to make it always positive and valid for all signals. The proposed definition of IF can easily be extended for multidimensional signals such as spatial data (image) and space-time data (3D wave equation).

A MATLAB implementation code of the proposed method is outlined in Algorithm~\ref{linoep_algo1}. In order to avoid unnecessary variations in the IF, before evaluating the IF, the mean and dominating low frequency component like trend present in the signal can be removed. These can be easily removed by any zero-phase low pass filtering operation~\cite{rslc9,rslc12}. Moreover, depending upon the requirements, signal can be decomposed into a set of desired frequency bands.

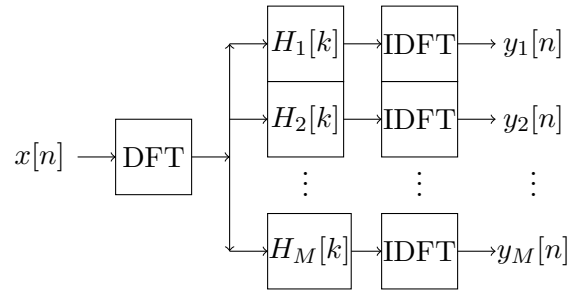
\begin{figure}[!t]
\centering
\begin{tikzpicture}
\node at (1,1.5) {$x[n]$};
\draw [black] (2,1) rectangle (3,2);
\node at (2.5,1.5) {DFT};

\draw [->] (1.5,1.5) -- (2,1.5);
\draw [->] (3,1.5) -- (3.5,1.5);
\draw [<->] (3.5,0.25) -- (3.5,0.5) -- (3.5,3);

\draw [black] (4,2.5) rectangle (5,3.5);
\draw [black] (4,1.5) rectangle (5,2.5);
\draw [black] (4,-.25) rectangle (5.1,.75);

\node at (4.5,1.25) {$\vdots$};

\draw [->] (3.5,3) -- (4,3);
\draw [->] (3.5,2) -- (4,2);
\draw [->] (3.5,0.25) -- (4,0.25);

\node at (4.5,3) {$H_1[k]$};
\node at (4.5,2) {$H_2[k]$};
\node at (4.55,0.25) {$H_M[k]$};

\draw [->] (5,3) -- (5.5,3);
\draw [->] (5,2) -- (5.5,2);
\draw [->] (5.1,0.25) -- (5.5,0.25);

\draw [black] (5.5,2.5) rectangle (6.5,3.5);
\draw [black] (5.5,1.5) rectangle (6.5,2.5);
\draw [black] (5.5,-.25) rectangle (6.5,.75);

\node at (6,3) {IDFT};
\node at (6,2) {IDFT};
\node at (6,1.25) {$\vdots$};
\node at (6,0.25) {IDFT};

\draw [->] (6.5,3) -- (7,3);
\draw [->] (6.5,2) -- (7,2);
\draw [->] (6.5,0.25) -- (7,0.25);

\node at (7.5,3) {$y_1[n]$};
\node at (7.5,2) {$y_2[n]$};
\node at (7.5,1.25) {$\vdots$};
\node at (7.5,0.25) {$y_M[n]$};
\end{tikzpicture}
\caption{Block diagram of the zero-phase filter-bank to decompose a signal $x[n]$ into a set $ \{ y_{1}[n], y_{2}[n], \cdots, y_{M}[n] \} $ of orthogonal desired frequency bands.}
\label{BlockDiag}
\end{figure}

In the examples presented in the following section, to decompose a signal $x[n]$ into a set of desired frequency bands, we have used the signal model
\begin{equation}
x[n]= c_0 + \displaystyle \sum_{i=1}^{M} y_{i}[n]=c_0 + \displaystyle \sum_{i=1}^{M} c_{i}[n], \label{sigmod1}
\end{equation}
where $c_0$ is the mean-value of signal $x[n]$, $\{y_{i}[n]\}_{i=1}^{M}$ and $\{c_{i}[n]\}_{i=1}^{M}$ are the $M$ orthogonal and LINOEP components, respectively.
 A simple block diagram of the zero-phase filter-bank (i.e. $H_i[k]\in \mathbb{R}$, $\forall i, k$) to decompose a signal $x[n]$ into a set $ \{ y_{1}[n], y_{2}[n], \cdots, y_{M}[n] \} $ is shown in Figure~\ref{BlockDiag}. The frequency response of $i$-th band of the DFT based zero-phase filter-bank can be defined by setting $H_i[k]=1$ at desired frequency band and zero otherwise, i.e.
\begin{equation}
 \left.\begin{aligned}
        H_i[k] & =1, \qquad (K_{i-1}+1) \le k \le K_i \text{ \& } \\
               &   \quad (N-K_{i}) \le k \le (N-K_{i-1}-1),\\
               & =0, \qquad \text{otherwise}
       \end{aligned}
 \right\} \label{Ch1_eqdft}
\end{equation}
where $i=1,2,\cdots, M$; $K_0=0$ and $K_M=N/2$ (or $K_M=(N-1)/2$ if $N$ is odd). Using the inverse DFT (IDFT), component $y_{i}[n]$ can be computed as
\begin{equation}
 y_{i}[n]=\displaystyle \sum_{k=0}^{N-1}\Big[H_i[k]X[k] \exp(j2\pi kn/N)\Big], \label{ylp}
\end{equation}
where $X[k]=\frac{1}{N}\displaystyle \sum_{n=0}^{N-1}x[n]\exp(-j2\pi kn/N)$ is the DFT of signal $x[n]$ of length $N$ samples.

%In this study, other than the DFT based zero-phase filter-bank (i.e. $H_i[k]=1$ at the desired frequencies and zero otherwise),
In this work, we use non-causal finite impulse response (FIR) and infinite impulse response (IIR) filter to decompose a signal into a set of LINOEP vectors by the filter mode decomposition (FMD) algorithm proposed in~\cite{rslc8,rslc9,rslc12}, as summarized in Algorithm~\ref{linoep_algo1}. In this algorithm, for each iteration, $ZPHPF_i$ ($ZPLPF_i$) is zero-phase high (low) pass filter (e.g. \emph{filtfilt} function of MATLAB) with desired cutoff frequency $f_{ci}$, and value of $\alpha_i$ is obtained such that $\mathbf{c}_i \perp \mathbf{\tilde{c}}_{i+1}$. It is to be noted that, in general, filter is not ideal (non brick wall frequency response) and therefore ${\mathbf{c}_i} \not\perp {\mathbf{c}_l}$ for ${i,l=1,2,\dots,M-1}$ and only ${\mathbf{c}_{M-1} \perp \mathbf{c}_{M}}$. We use PART A or PART B of algorithm to obtain $\{c_1,\cdots,c_M\}$ in order of highest to lowest or lowest to highest frequency components. The FMD with proposed IF can easily be adapted for multichannel and multidimensional data decomposition into a set of AM-FM components.

We advocate to use zero-phase filtering because it preserves salient features such as minima and maxima in the filtered waveform exactly at the position where those features occur in the unfiltered waveform. It is pertinent to note that the conventional (i.e. non zero-phase) filtering shifts these features in the filtered waveform and therefore cannot be used to obtain a meaningful TFE distribution, which is clearly demonstrated in simulation results. The zero-phase filtering of a signal can be obtained by the DFT, non-causal FIR and IIR filters, or via other decomposition methods like FDM, VMD, EVD, EWT and EMD algorithms.
\begin{algorithm}[!t]
%function [freq amp]=timefrequency_estimation(x,Fs)
  %{x}={data}-mean({data});
  {z}=hilbert({x});{amp}=abs({z});{phi}=unwrap(angle({z}));
  % if diffPhase is -ve, make it +ve by adding pi\;
  diffPhase=diff(phi);index=find(diffPhase$<$0)\;
  diffPhase(index)=diffPhase(index)+pi;
  frequncy=[diffPhase;diffPhase(end)]*(Fs/(2*pi));
  %frequency=diffPhase*(Fs/(2*pi));

 \caption{A MATLAB implementation code of the proposed method to estimate amplitude and frequency of data $x$. Here, Fs is the sampling frequency.}\label{linoep_algo1}
\end{algorithm}\label{linoep_algo0}
\begin{algorithm}[!t]
\%PART A\\
 $\mathbf{x}_{1}=\mathbf{x}$\;
 \For{$i=1$ to $M-1$}
 {
  %$\mathbf{y}_i=ZeroPhaseHighPassFilter_i(\mathbf{x}_i,f_{ci})$\;
  $\mathbf{y}_i=ZPHPF_i(\mathbf{x}_i,f_{ci})$\;
  $\mathbf{r}_i=\mathbf{x}_i-\mathbf{y}_i$\;
  ${\alpha_i=\frac{\langle \mathbf{y}_i,\mathbf{r}_i \rangle }{\langle \mathbf{r}_i,\mathbf{r}_i \rangle }}$\;
  $\mathbf{c}_i= \mathbf{y}_i- \alpha_i \mathbf{r}_i$\;
  $\mathbf{\tilde{c}}_{i+1}= (1+\alpha_i) \mathbf{r}_i$\;
  $\mathbf{x}_{i+1}=\mathbf{\tilde{c}}_{i+1}$\;
 }
 $\mathbf{c}_{M}=\mathbf{\tilde{c}}_{M}$\;
 \%PART B\\
$\mathbf{x}_{1}=\mathbf{x}$\;
 \For{$i=1$ to $M-1$}
 {
  %$\mathbf{y}_i=ZeroPhaseHighPassFilter_i(\mathbf{x}_i,f_{ci})$\;
  $\mathbf{y}_i=ZPLPF_i(\mathbf{x}_i,f_{ci})$\;
  $\mathbf{r}_i=\mathbf{x}_i-\mathbf{y}_i$\;
  ${\alpha_i=\frac{\langle \mathbf{r}_i,\mathbf{y}_i \rangle }{\langle \mathbf{y}_i,\mathbf{y}_i \rangle }}$\;
  $\mathbf{c}_i= (1+\alpha_i) \mathbf{y}_i $\;
  $\mathbf{\tilde{c}}_{i+1}= \mathbf{r}_i- \alpha_i \mathbf{y}_i$\;
  $\mathbf{x}_{i+1}=\mathbf{\tilde{c}}_{i+1}$\;
 }
 $\mathbf{c}_{M}=\mathbf{\tilde{c}}_{M}$\;
 \caption{An FMD algorithm to obtain LINOEP vectors $\mathbf{c}_i$ from decomposition of a signal $\mathbf{x}$ such that $\mathbf{x}=c_0+\sum_{i=1}^{M}\mathbf{c}_i$ and ${\mathbf{c}_i \perp \sum_{l=i+1}^{M}\mathbf{c}_l}$. Use PART A (PART B) to obtain $\{c_1,\cdots,c_M\}$ in order of highest to lowest (lowest to highest) frequency components.% or use PART B to obtain $\{c_1,\cdots,c_M\}$ in order of lowest to highest frequency components
 } \label{linoep_algo1}
\end{algorithm}
%\begin{algorithm}[!t]
% $\mathbf{x}_{1}=\mathbf{x}$\;
% \For{$i=1$ to $M-1$}
% {
%  $\mathbf{y}_i=ZPLPF_i(\mathbf{x}_i,f_{ci})$\;
%  $\mathbf{r}_i=\mathbf{x}_i-\mathbf{y}_i$\;
%  ${\alpha_i=\frac{\langle \mathbf{r}_i,\mathbf{y}_i \rangle }{\langle \mathbf{y}_i,\mathbf{y}_i \rangle }}$\;
%  $\mathbf{c}_i= (1+\alpha_i) \mathbf{y}_i $\;
%  $\mathbf{\tilde{c}}_{i+1}= \mathbf{r}_i- \alpha_i \mathbf{y}_i$\;
%  $\mathbf{x}_{i+1}=\mathbf{\tilde{c}}_{i+1}$\;
% }
% $\mathbf{c}_{M}=\mathbf{\tilde{c}}_{M}$\;
% \caption{An FMD algorithm to obtain LINOEP vectors $\mathbf{c}_i$ from decomposition of a signal $\mathbf{x}$ such that $\mathbf{x}=c_0+\sum_{i=1}^{M}\mathbf{c}_i$ and ${\mathbf{c}_i \perp \sum_{l=i+1}^{M}\mathbf{c}_l}$, where $c_1,\cdots,c_M$ are in order of lowest to highest frequency components.}\label{linoep_algo2}
%\end{algorithm}
\section{SIMULATION RESULTS AND DISCUSSION}\label{simre}
In this section, we consider number of examples that are mostly discussed in literature to validate the efficacy of method understudy. A complete MATLAB code of the proposed method is publicly available for download at~\cite{MTCode}.

\begin{figure}[!t]
\centering
\includegraphics[angle=0,width=0.5\textwidth,height=0.3\textwidth]{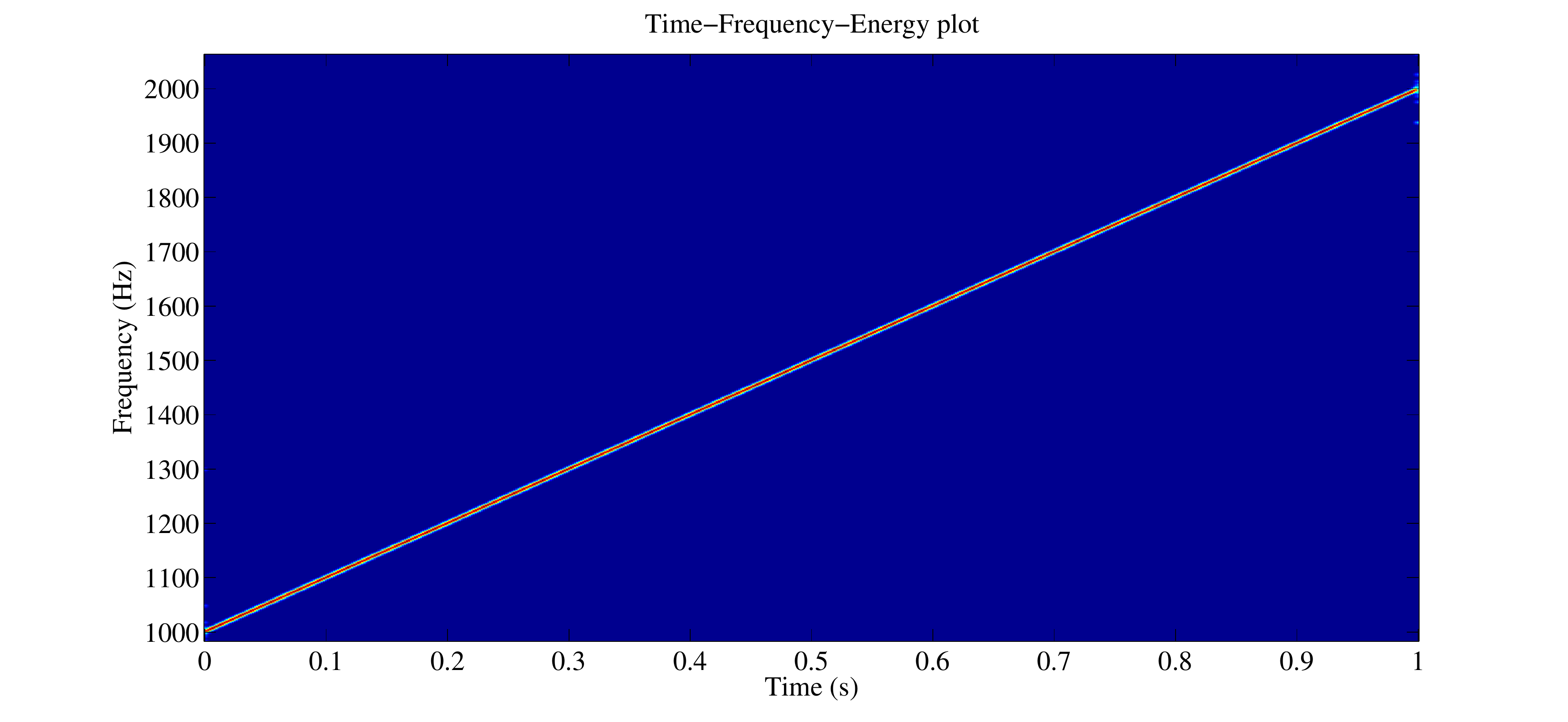}
\includegraphics[angle=0,width=0.5\textwidth,height=0.3\textwidth]{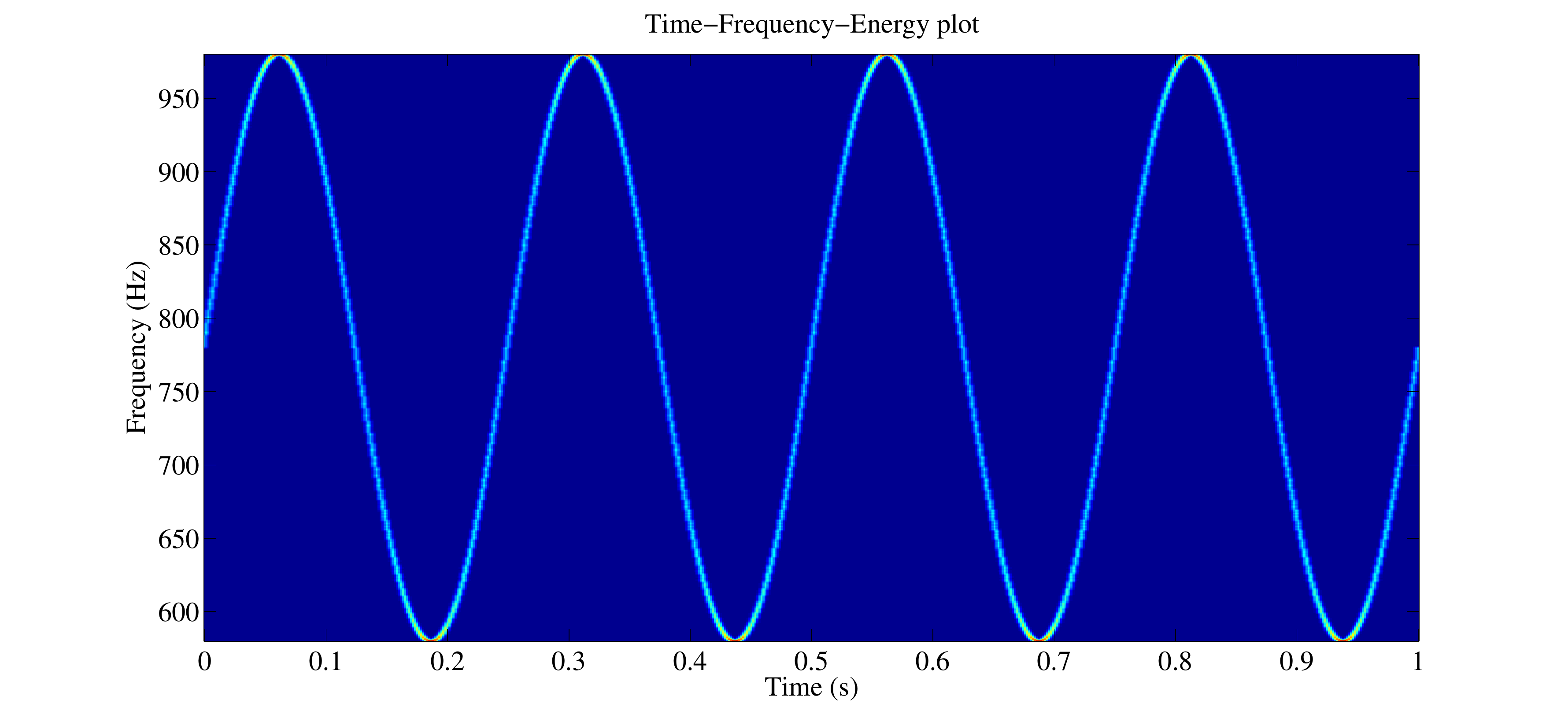}
\includegraphics[angle=0,width=0.5\textwidth,height=0.3\textwidth]{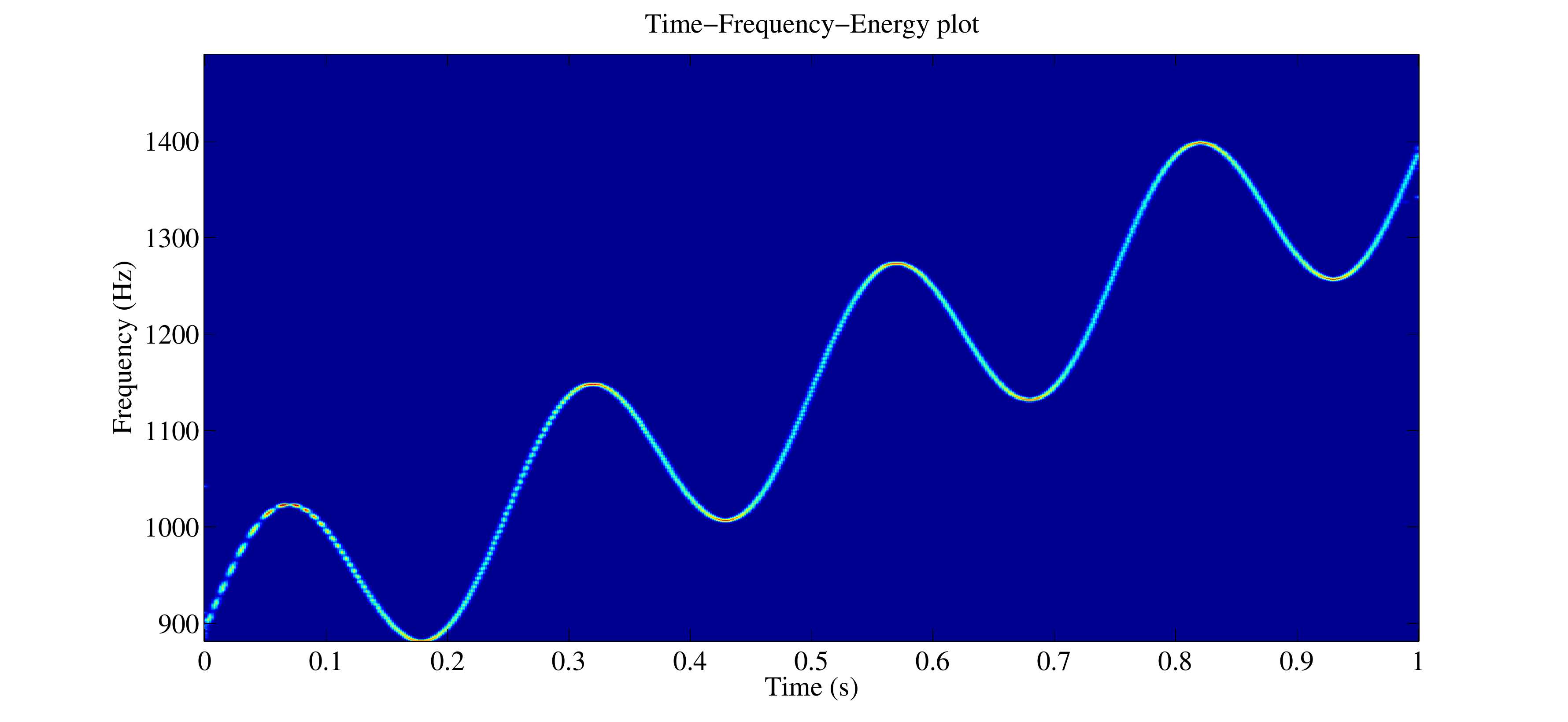}
\caption{The TFE analysis of nonstationary signals by this proposed method without decomposition: (top figure) linear chirp, (middle figure) frequency modulated signal and (bottom figure) sum of linear chirp and
FM signals.}
\label{fig:lcPlusFmp}
\end{figure}

\textbf{Example 1:} In this example, we consider three nonstationary signals (a) linear chirp (unit amplitude and [1000--2000] Hz), (b) frequency modulated (FM) sinusoid (unit amplitude, carrier $f_c=780$ Hz and frequency deviation 200 Hz) and (c) mixture of a linear chirp and frequency modulated (FM) signals, i.e sum of signals of cases (a) and (b). Figure~\ref{fig:lcPlusFmp} (top to bottom) shows the time-frequency-energy (TFE) estimates of these three nonstationary signals [(a), (b) and (c)] obtained by using the proposed IF without decomposition. We observe that the frequencies in case (c), Figure~\ref{fig:lcPlusFmp} (bottom one), are average frequencies of the first two cases (a) and (b).

\begin{figure}[!t]
\centering
\includegraphics[angle=0,width=0.5\textwidth,height=0.3\textwidth]{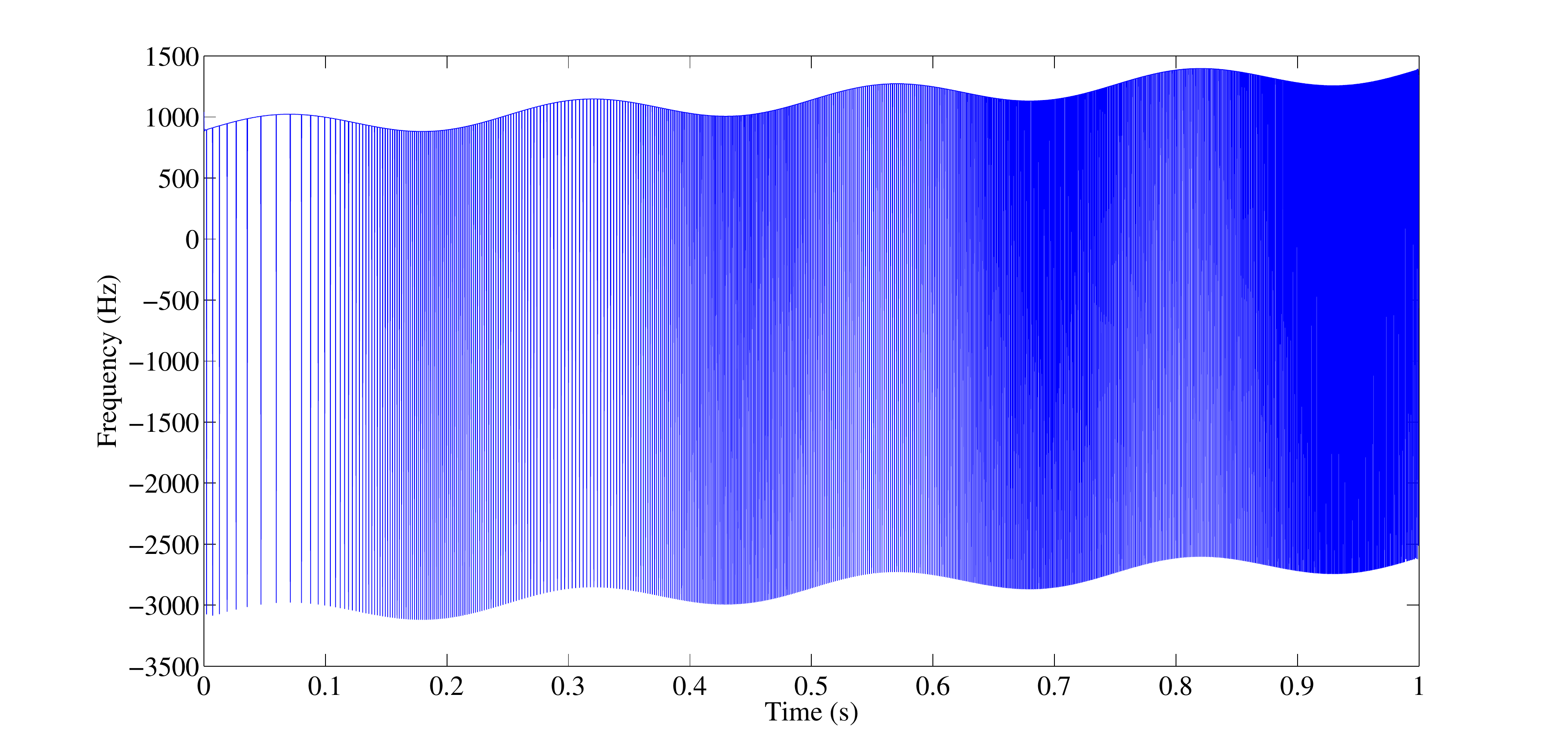}
\includegraphics[angle=0,width=0.5\textwidth,height=0.3\textwidth]{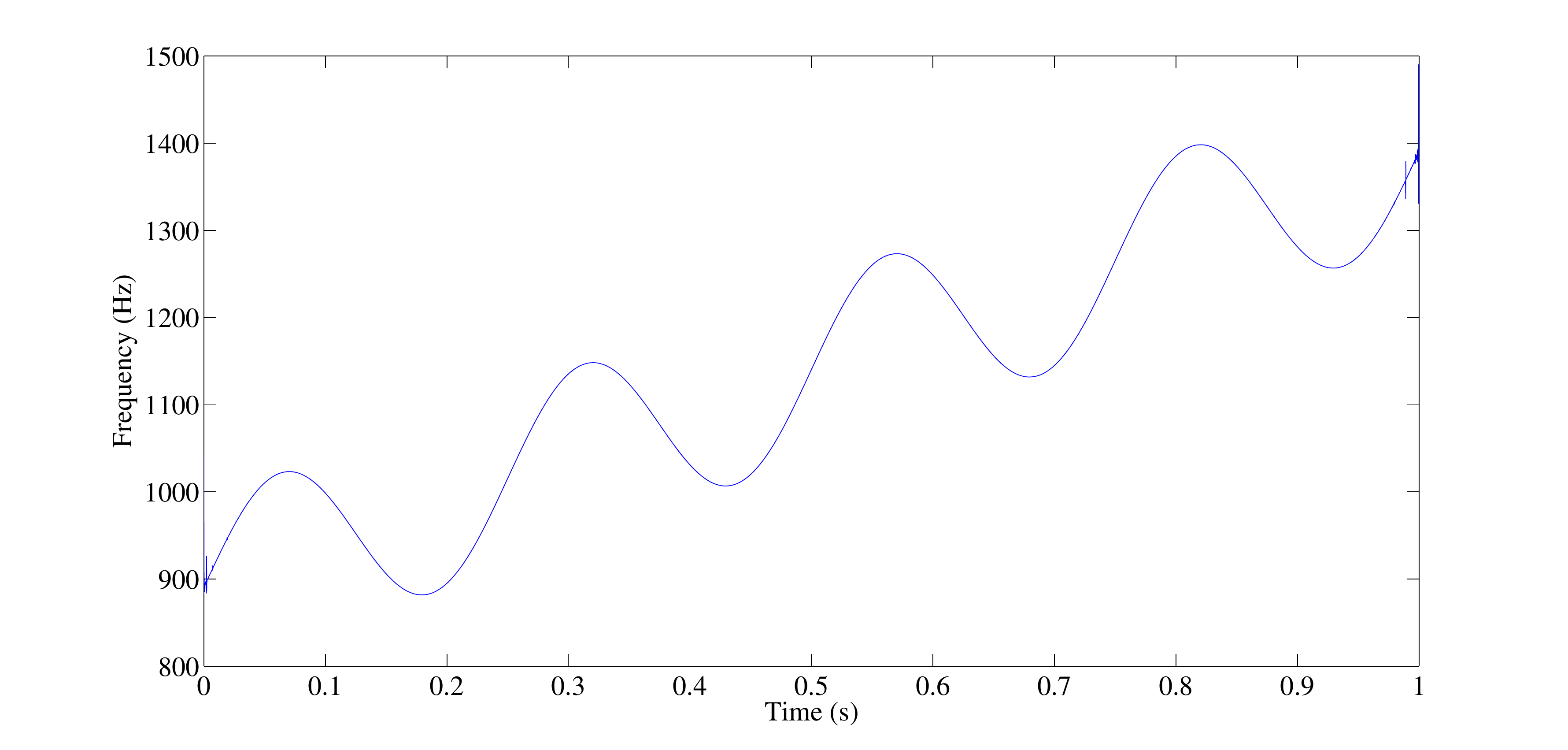}
\caption{The instantaneous frequency (IF) estimates of nonstationary signal, which is sum of linear chirp and FM signals, without decomposition: (top figure) with conventional IF~\eqref{sub5} and (bottom figure) with proposed IF~\eqref{Ch1_eq12}. Notice the both positive and negative frequencies with conventional IF (top figure); only positive and correct values of frequencies by the proposed IF (bottom figure).}
\label{fig:lcPlusFmp0}
\end{figure}

Figure~\ref{fig:lcPlusFmp0} shows the IF estimates of nonstationary signal, which is sum of linear chirp and FM signals i.e. case (c), without decomposition: (1) top figure with conventional IF~\eqref{sub5} where notice both the positive and negative estimates of frequencies, (2) bottom figure with proposed IF~\eqref{Ch1_eq12} that produces only positive and correct values of frequencies. This, clearly, demonstrate that the proposed definition of the IF is able to obtain correct values of frequencies at all times.

\begin{figure}[!t]
\centering
\includegraphics[angle=0,width=0.5\textwidth,height=0.3\textwidth]{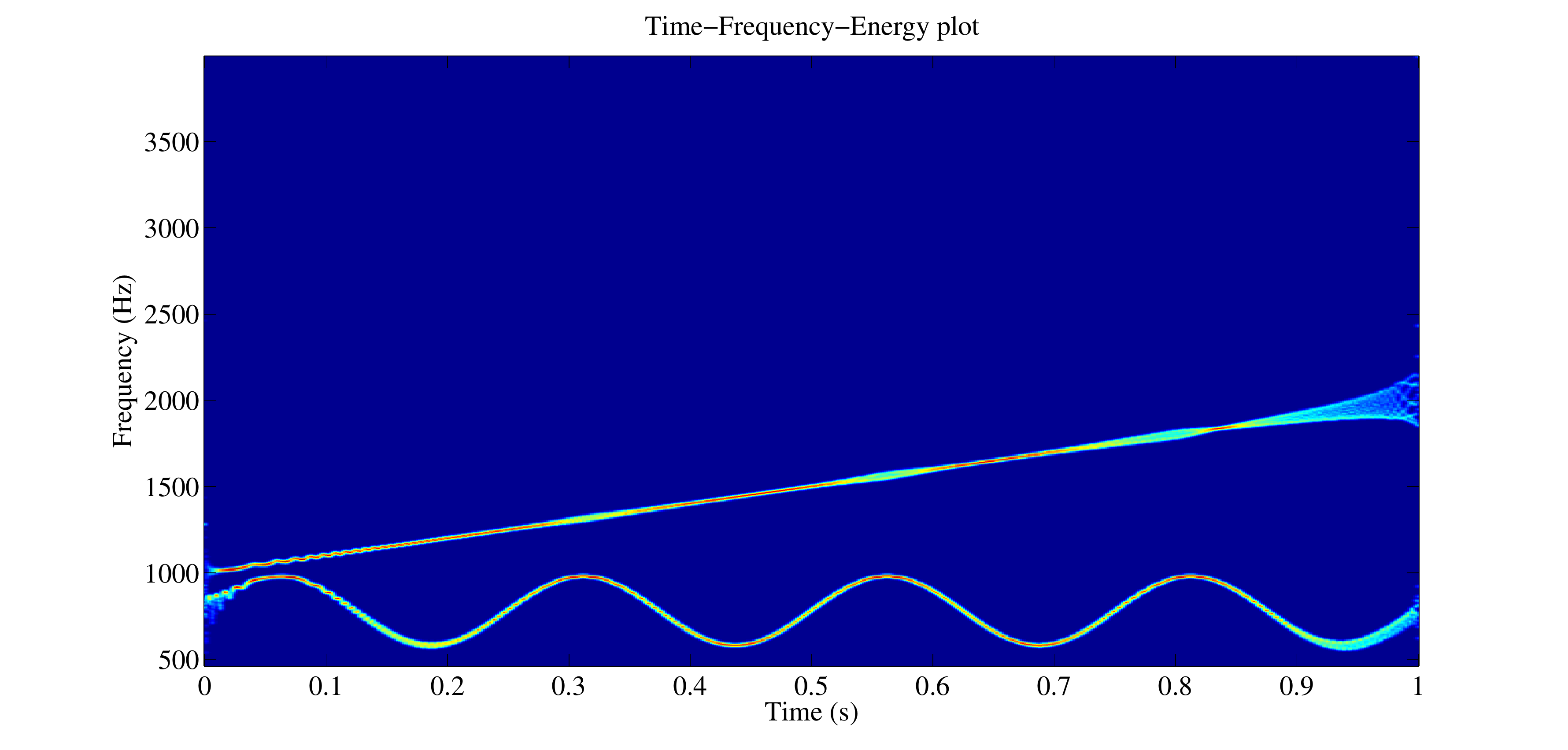}
\includegraphics[angle=0,width=0.5\textwidth,height=0.3\textwidth]{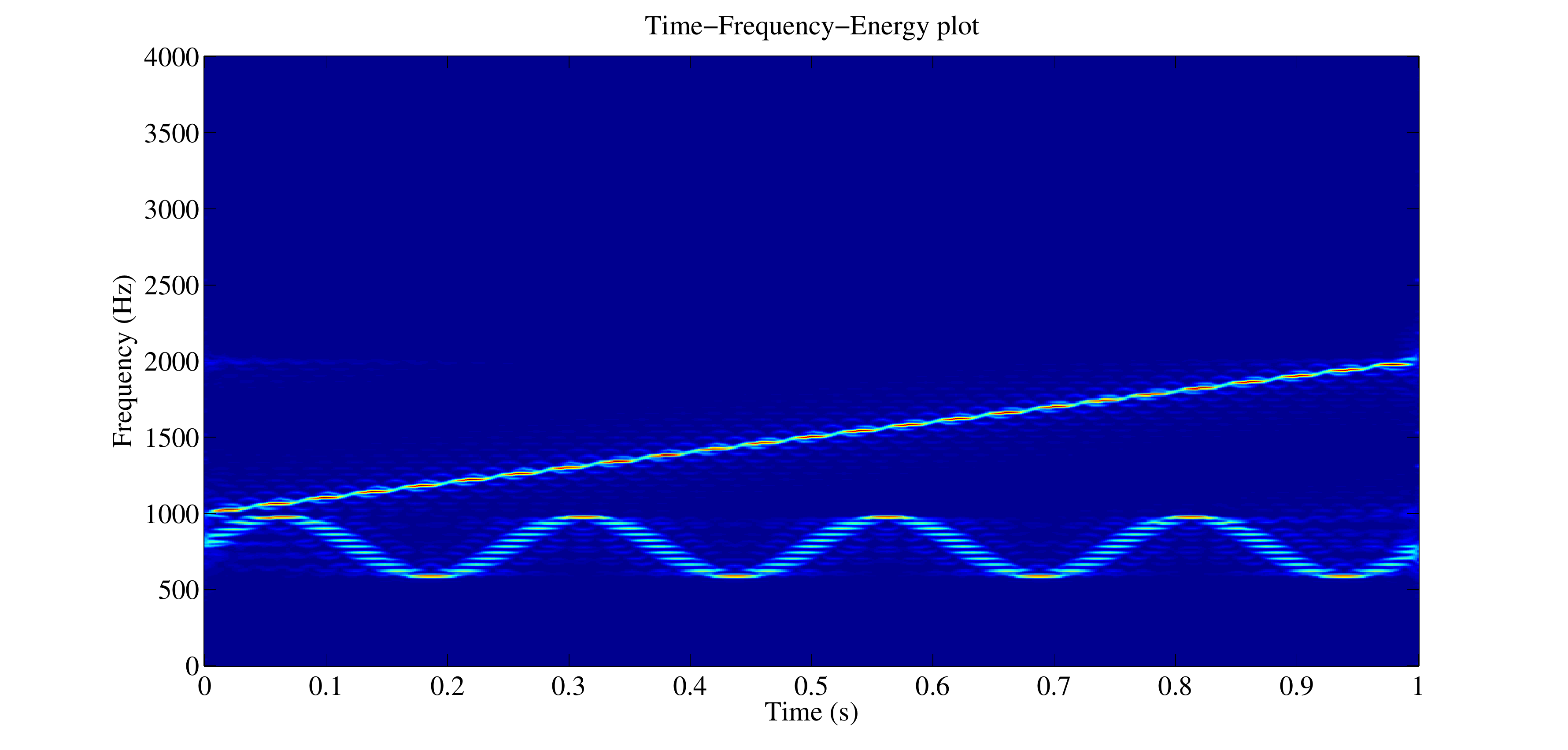}
\caption{The TFE analysis of nonstationary signal, which is sum of linear chirp and FM signals, by this proposed method with DFT based decomposition: (top figure) into 2 bands of [0--1000, 1000--4000] Hz and (bottom figure) into 100 bands of equal bandwidth.}
\label{fig:lcPlusFmp1}
\end{figure}

Figure~\ref{fig:lcPlusFmp1} shows the TFE estimates of nonstationary signal, case (c), by this proposed method with decomposition: (top figure) into 2 bands of [0--1000, 1000--4000] Hz and (bottom figure) into 100 bands of equal bandwidth of 40 Hz (which is $F_{max}/100$ where $F_{max}$ is the highest frequency component present in the signal and it is equal to half of the sampling frequency i.e., $F_{max}=F_s/2$). The proposed method is, clearly, able to track TFE distribution present in all these signals.

\begin{figure}[!t]
\centering
\includegraphics[angle=0,width=0.5\textwidth,height=0.3\textwidth]{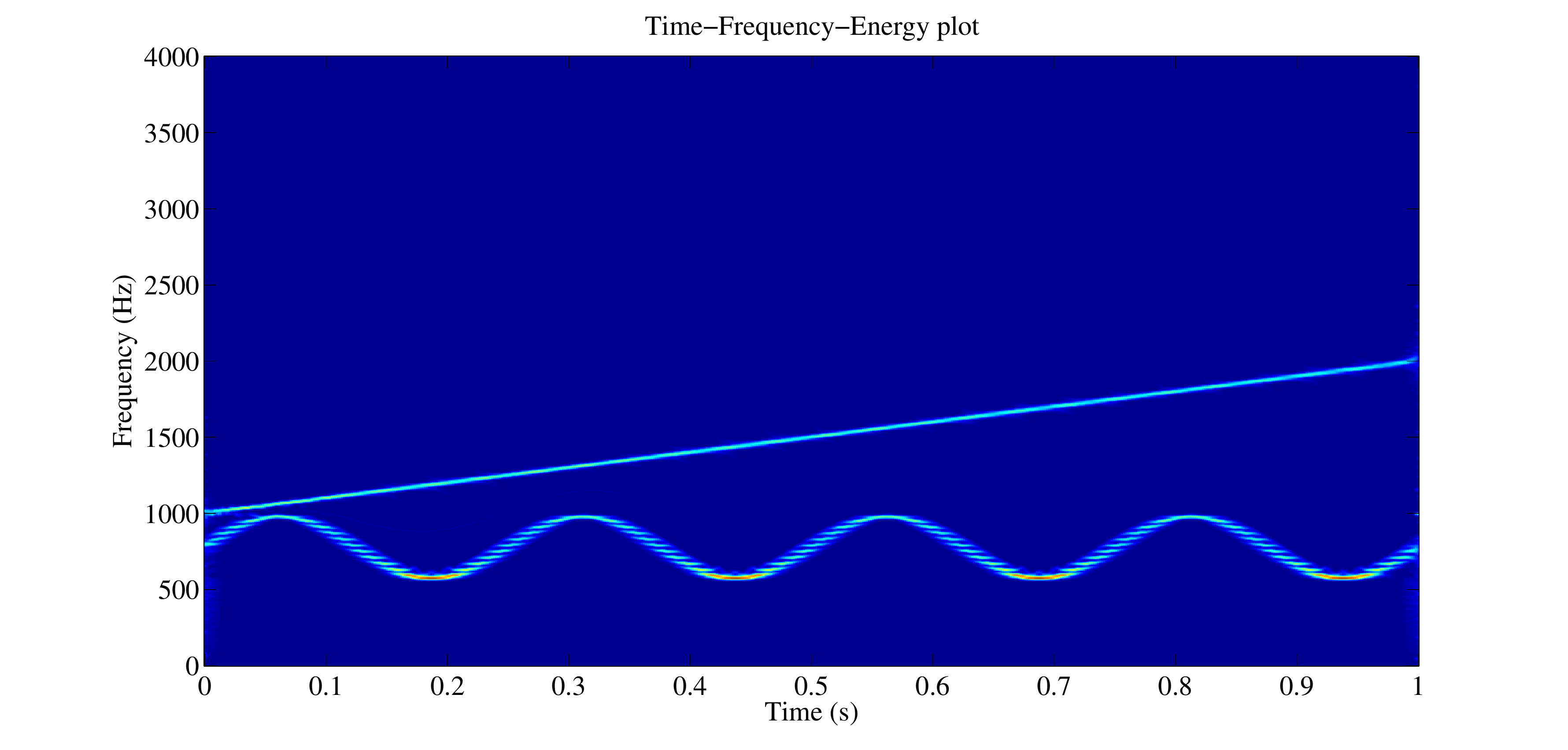}
\includegraphics[angle=0,width=0.5\textwidth,height=0.3\textwidth]{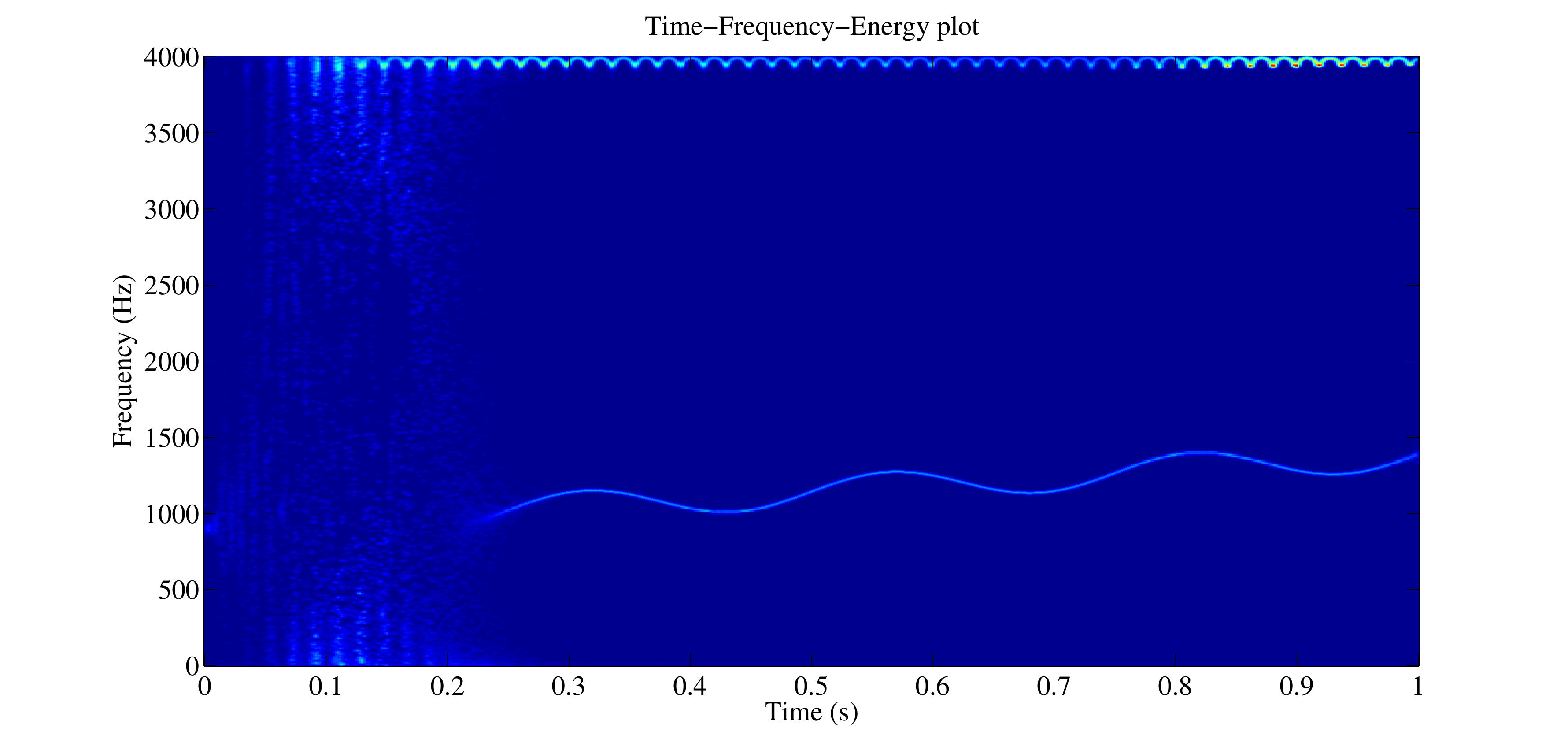}
\caption{The TFE analysis of nonstationary signal, which is sum of linear chirp and FM signals, by this proposed method: (top figure) with zero-phase FIR filter based decomposition into 100 bands and (bottom figure) with conventional (non zero-phase) FIR filter based decomposition into 100 bands of equal bandwidth.}
\label{fig:lcPlusFmp2fmd}
\end{figure}

Figure~\ref{fig:lcPlusFmp2fmd} shows the TFE estimates of nonstationary signal, case (c), by this proposed IF: (top figure) with zero-phase FIR filter based decomposition into 100 bands; and (bottom figure) with conventional (non zero-phase) FIR filter based decomposition into 100 bands of equal bandwidth, which is not able to detect and track true frequencies present in the signal.

\begin{figure}[!t]
\centering
\includegraphics[angle=0,width=0.5\textwidth,height=0.3\textwidth]{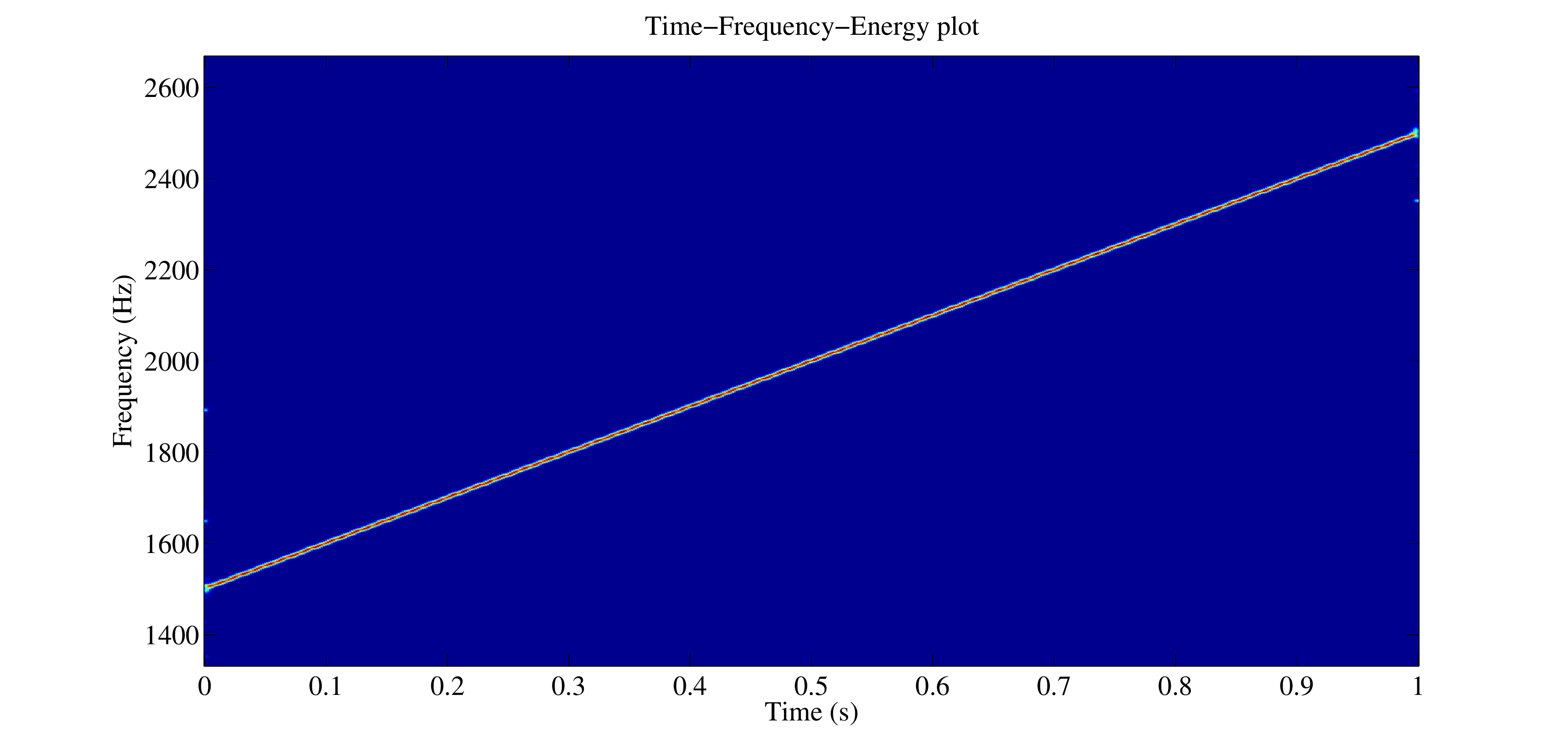}
\includegraphics[angle=0,width=0.5\textwidth,height=0.3\textwidth]{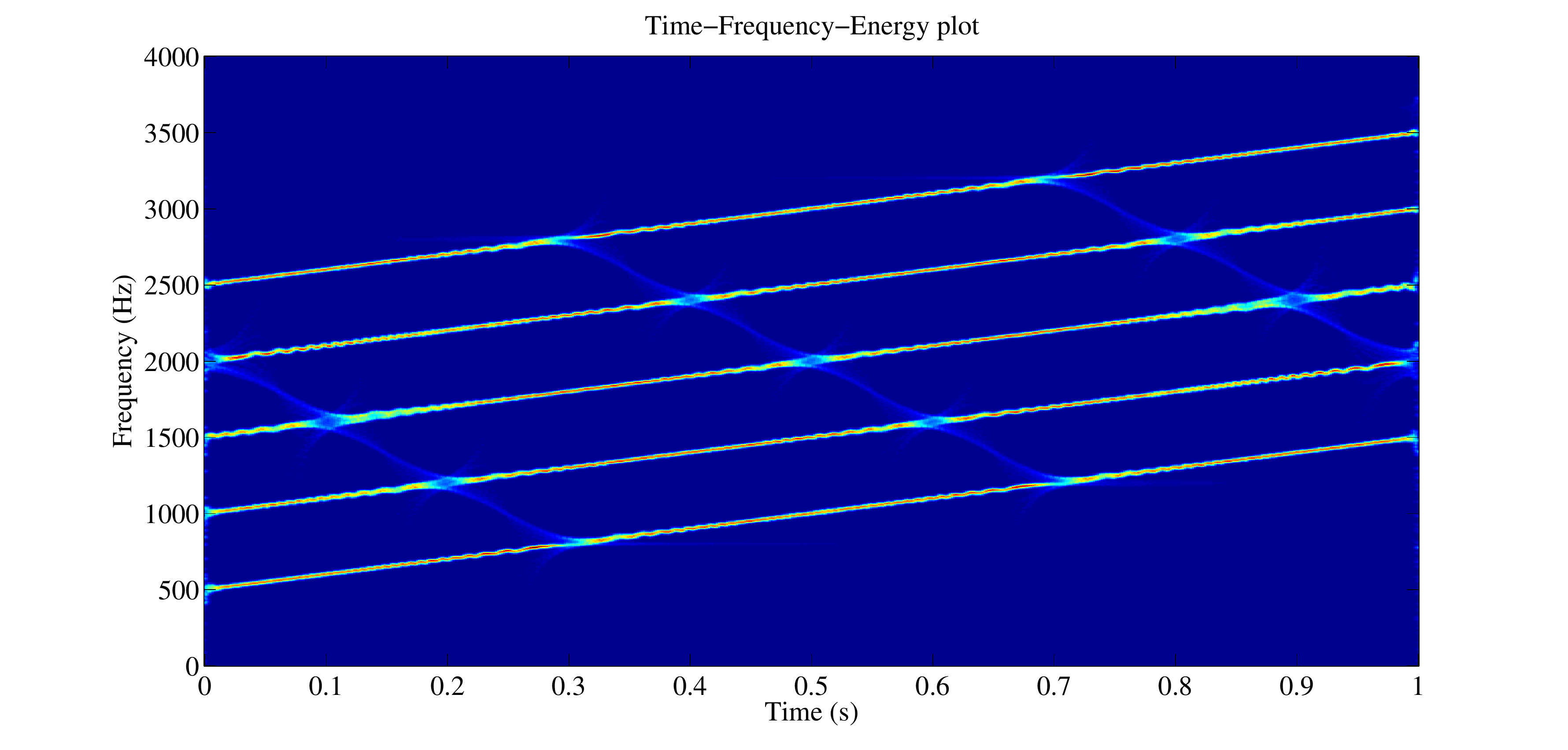}
\includegraphics[angle=0,width=0.5\textwidth,height=0.3\textwidth]{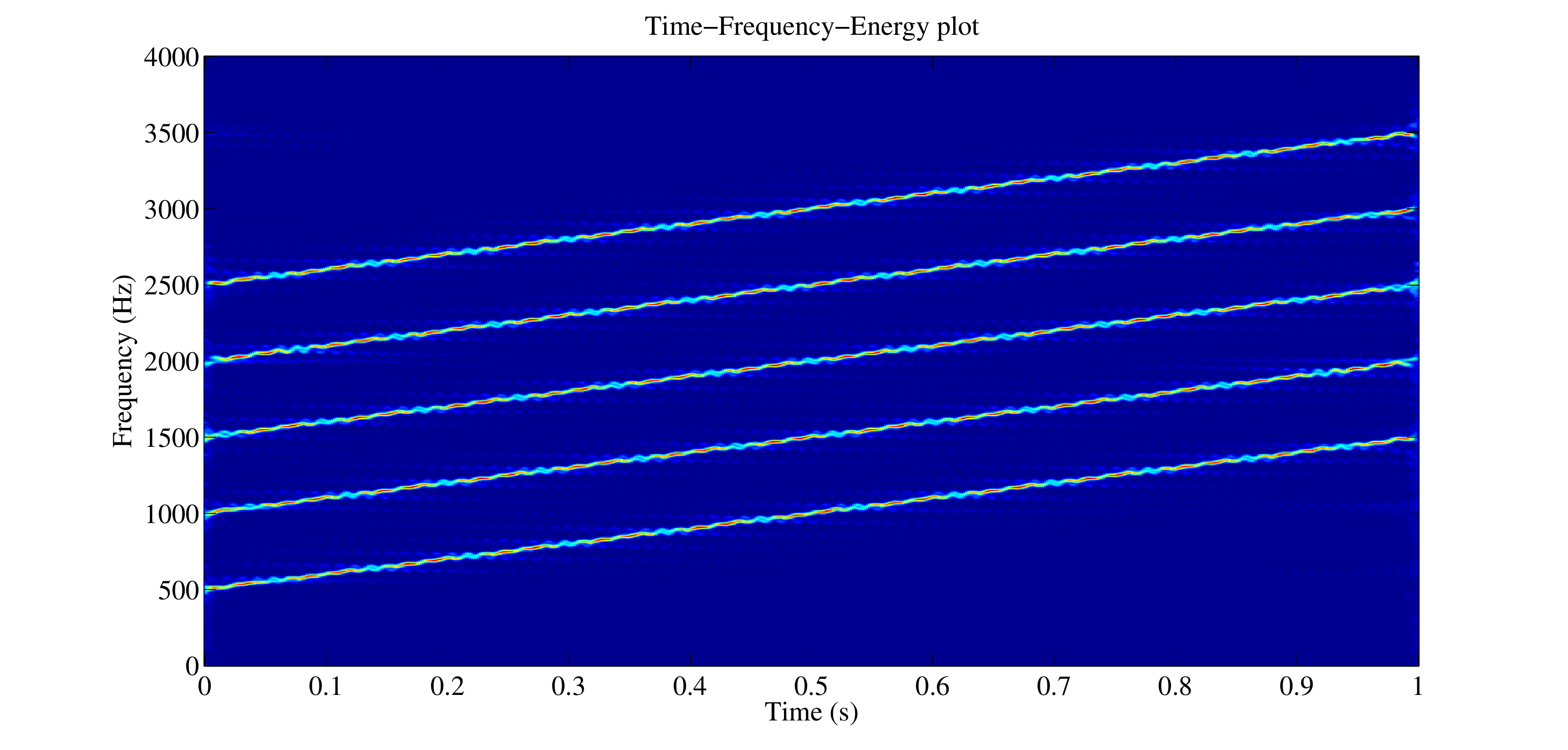}
\caption{The TFE analysis of a nonstationary signal, which is sum of five linear chirp signals, by this proposed method: (top figure) without decomposition, (middle figure) with DFT based decomposition into 10 bands and (bottom figure) with DFT based decomposition into 20 bands of equal bandwidth.}
\label{fig:ParallelChirps}
\end{figure}

\textbf{Example 2:} We obtain a nonstationary signal by adding five unit amplitude linear chirps of frequencies [500--1500] Hz, [1000--2000] Hz, [1500--2500] Hz, [2000--3000] Hz and [2500--3500] Hz. Figure~\ref{fig:ParallelChirps} shows the TFE analysis of this nonstationary signal, which is sum of five linear chirp signals, by this proposed IF: (top figure) without decomposition that presents average frequencies [1500--2500] Hz, which are average of frequencies present in five chirp signals; (middle figure) with DFT based decomposition into 10 bands and (bottom figure) with DFT based decomposition into 20 bands of equal bandwidth. These two (middle and bottom one) figures clearly reveal the five chirp signals present in the signal under analysis.

\begin{figure}[!t]
\centering
\includegraphics[angle=0,width=0.5\textwidth,height=0.3\textwidth]{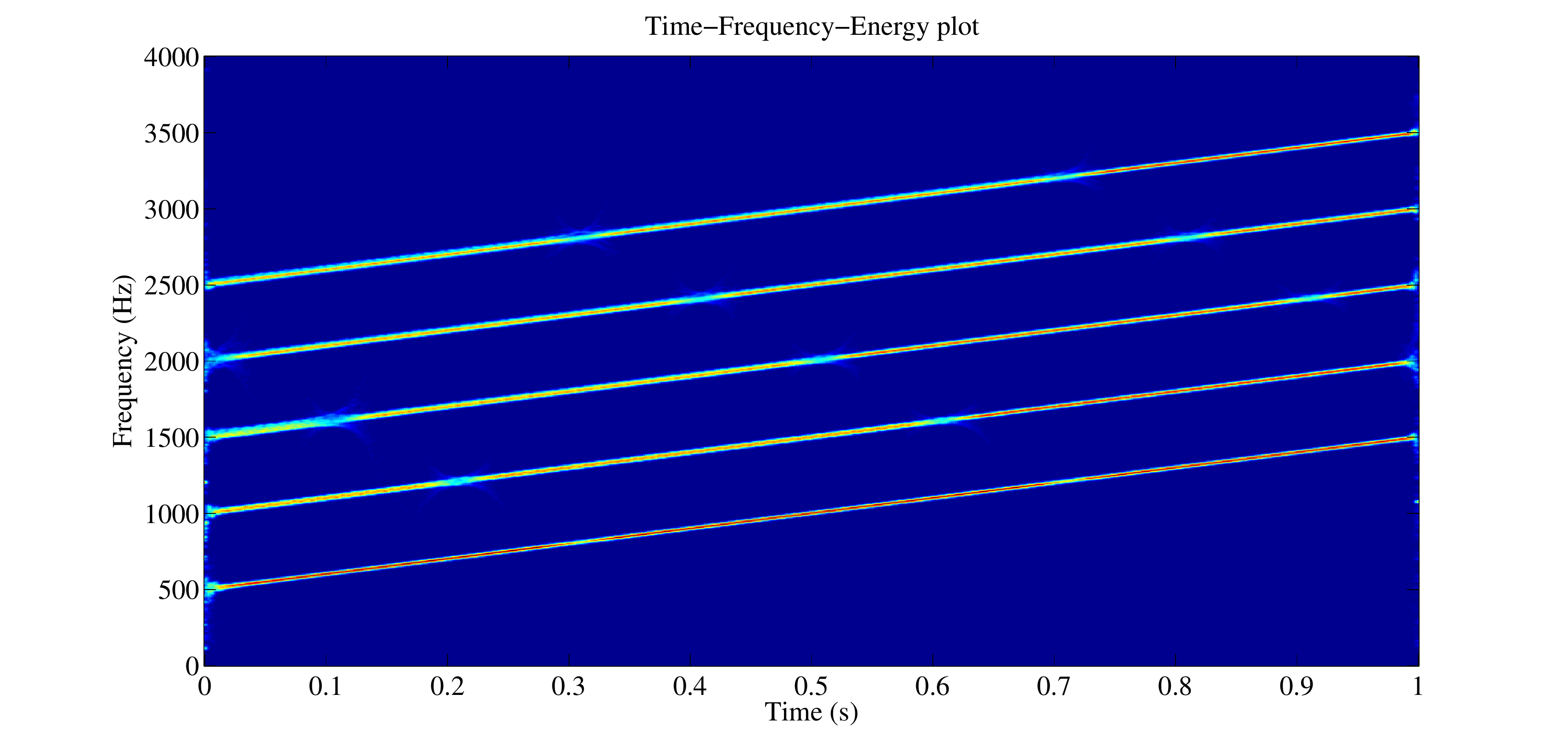}
\includegraphics[angle=0,width=0.5\textwidth,height=0.3\textwidth]{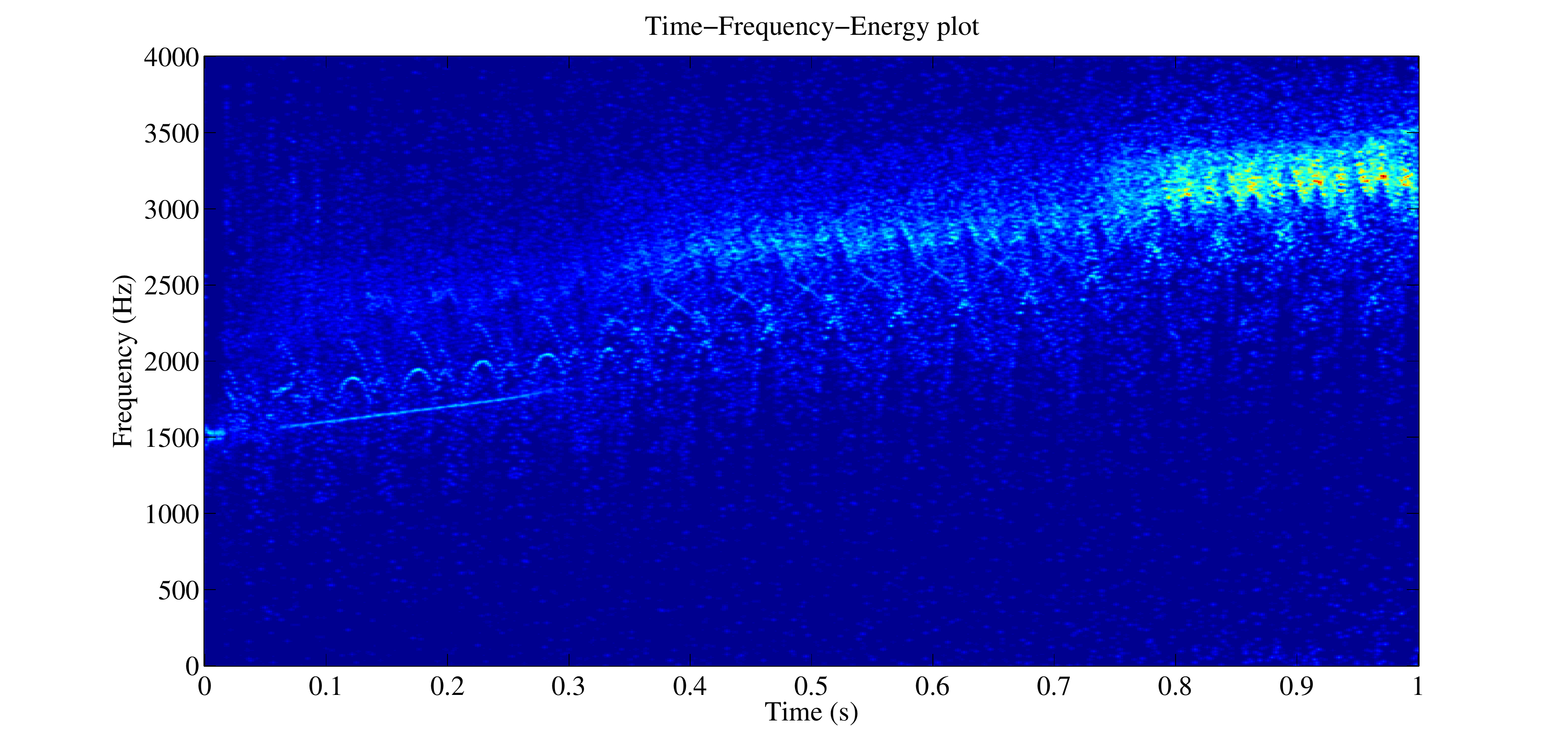}
\includegraphics[angle=0,width=0.5\textwidth,height=0.3\textwidth]{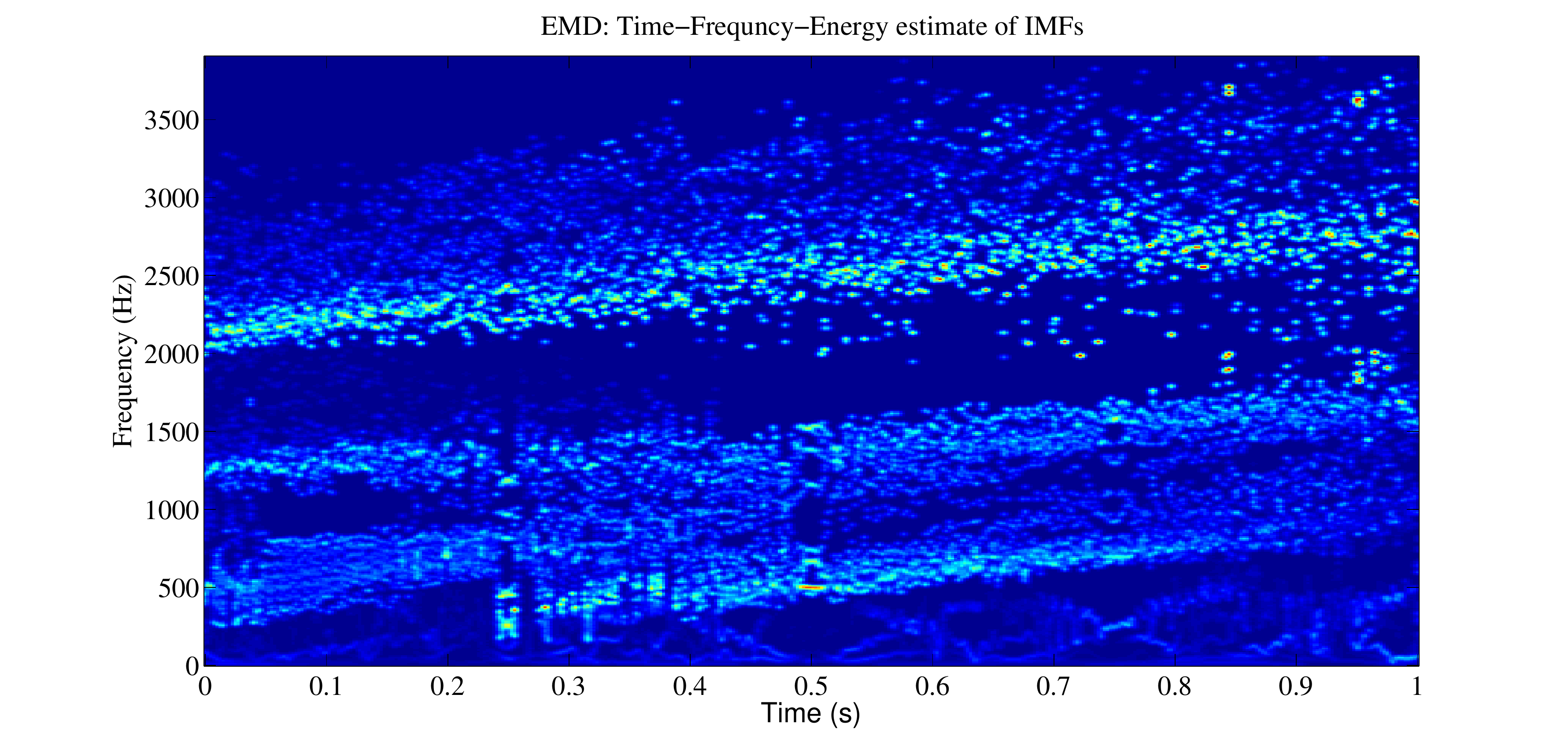}
\caption{The TFE analysis of a nonstationary signal, which is sum of five linear chirp signals (1) by this proposed method with zero-phase FIR filter based decomposition into 10 bands (upper figure), and with conventional (non zero-phase) FIR filter based decomposition into 10 bands of equal bandwidth (middle figure); (2) by the EMD algorithm (lower figure).}
\label{fig:ParallelChirps1FIR}
\end{figure}

Figure~\ref{fig:ParallelChirps1FIR} shows the TFE distribution of a nonstationary signal (sum of five linear chirp signals) (1) by this proposed method with zero-phase FIR filter based decomposition into 10 bands (upper figure), and with conventional (non zero-phase) FIR filter based decomposition into 10 bands of equal bandwidth (middle figure); and (2) by the EMD algorithm (lower figure). Clearly, both the conventional and EMD algorithm not able to detect true frequencies present in the signal.

\begin{figure}[!t]
\centering
\includegraphics[angle=0,width=0.5\textwidth,height=0.3\textwidth]{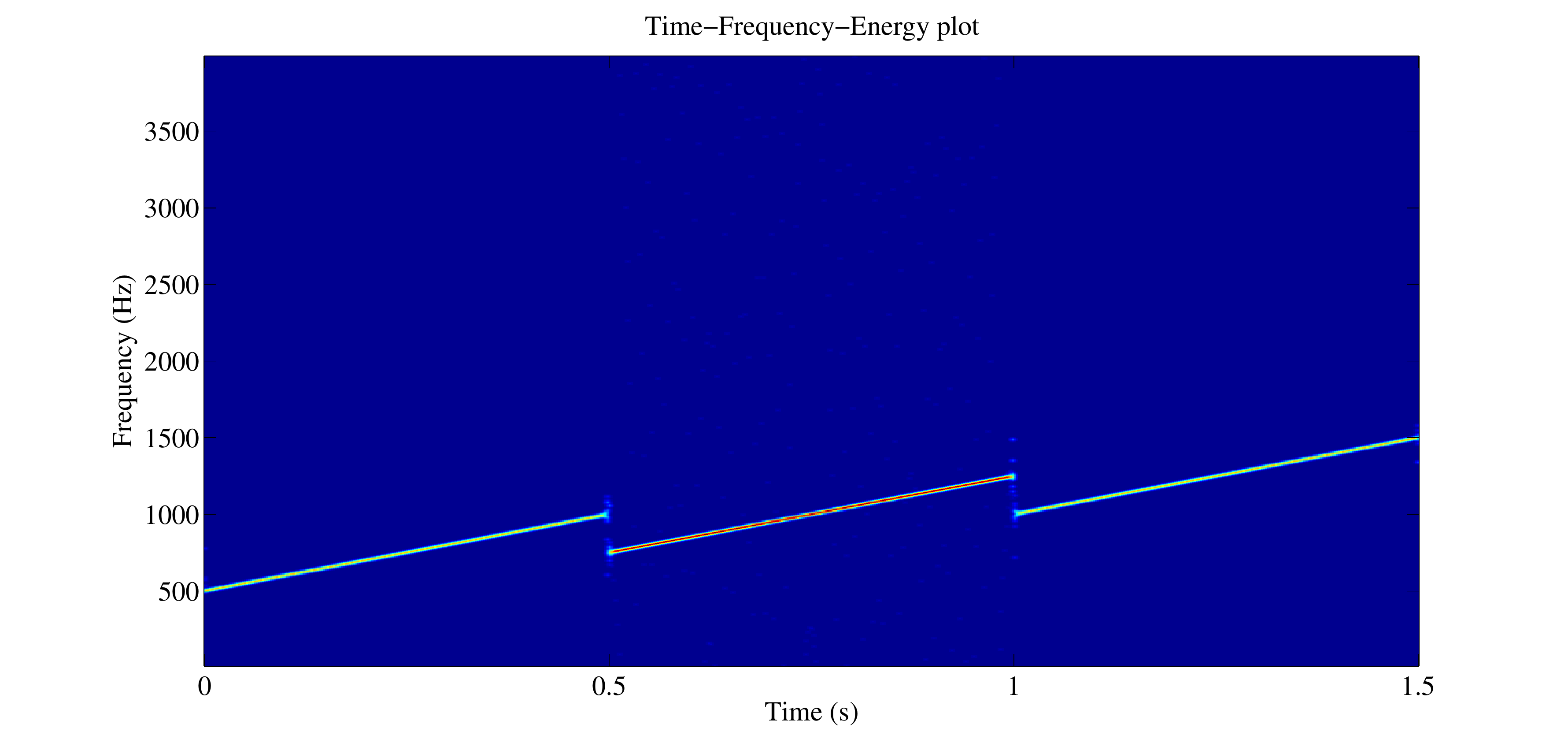}
\includegraphics[angle=0,width=0.5\textwidth,height=0.3\textwidth]{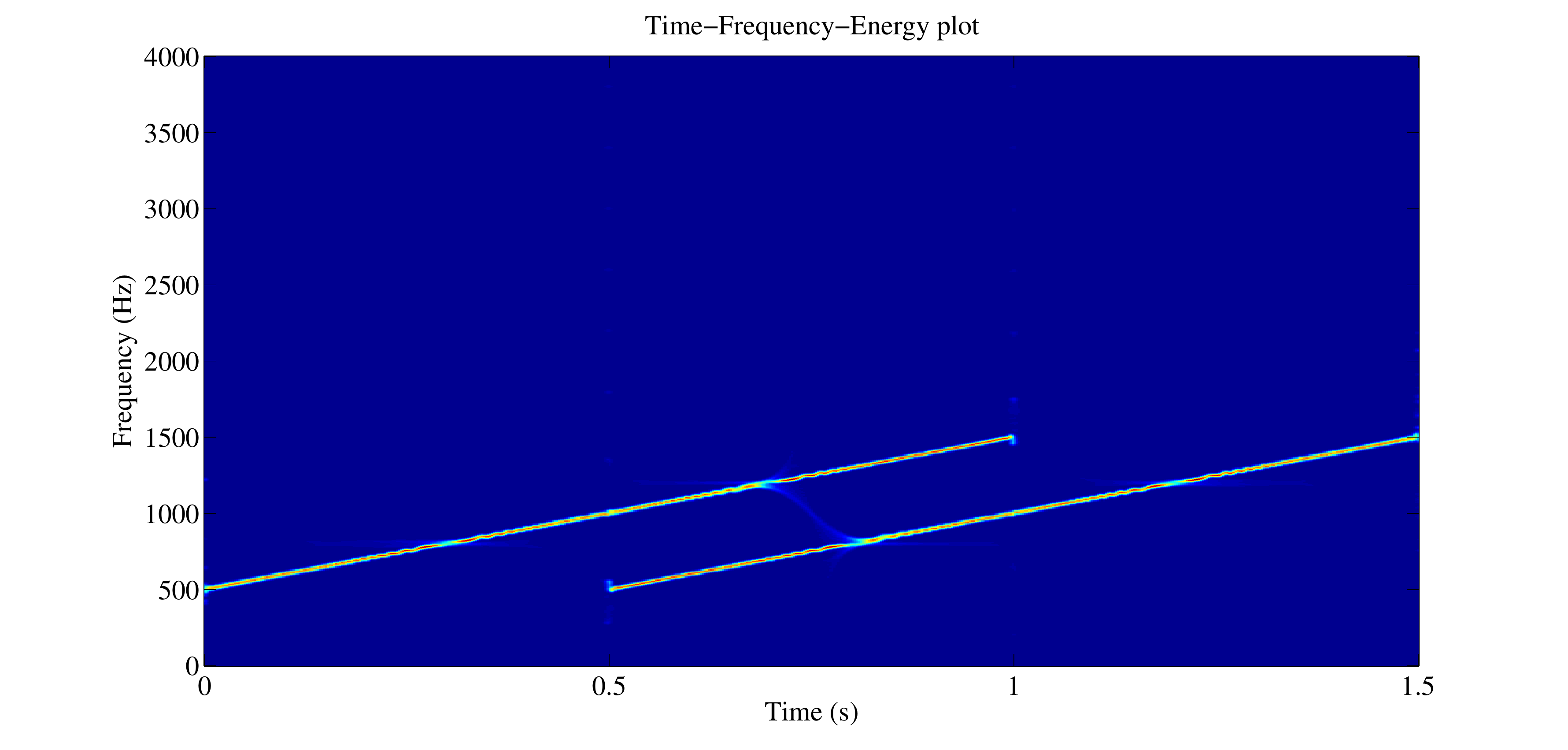}
\caption{The TFE analysis of a nonstationary signal, which is sum of a linear chirp signal and its delayed version, by this proposed method: (top figure) without decomposition, and (bottom figure) with DFT based decomposition into 10 bands of equal bandwidth.}
\label{fig:ChirpAndItsDelayed}
\end{figure}

\textbf{Example 3:} In Figure~\ref{fig:ChirpAndItsDelayed}, we present the TFE analysis of a nonstationary signal, which is sum of a unit amplitude linear chirp signal [500-1500] Hz ([0--1] s chirp signal and [1--1.5] s no signal) and its delayed version ([0--0.5] s no signal and [0.5--1.5] s same chirp signal), by this proposed method. The top figure is obtained without decomposition that has three parts: (1) [0--0.5] s chirp signal frequencies, (2) [0.5--1] s average of overlapped chip signals frequencies and (3) [1--1.5] s chirp signal frequencies of delayed version. The bottom figure is obtained by DFT based decomposition of signal into 10 bands of equal bandwidth, which clearly reveals that the signal understudy is sum of a chirp signal and its delayed version.

\begin{figure}[!t]
\centering
\includegraphics[angle=0,width=0.5\textwidth,height=0.3\textwidth]{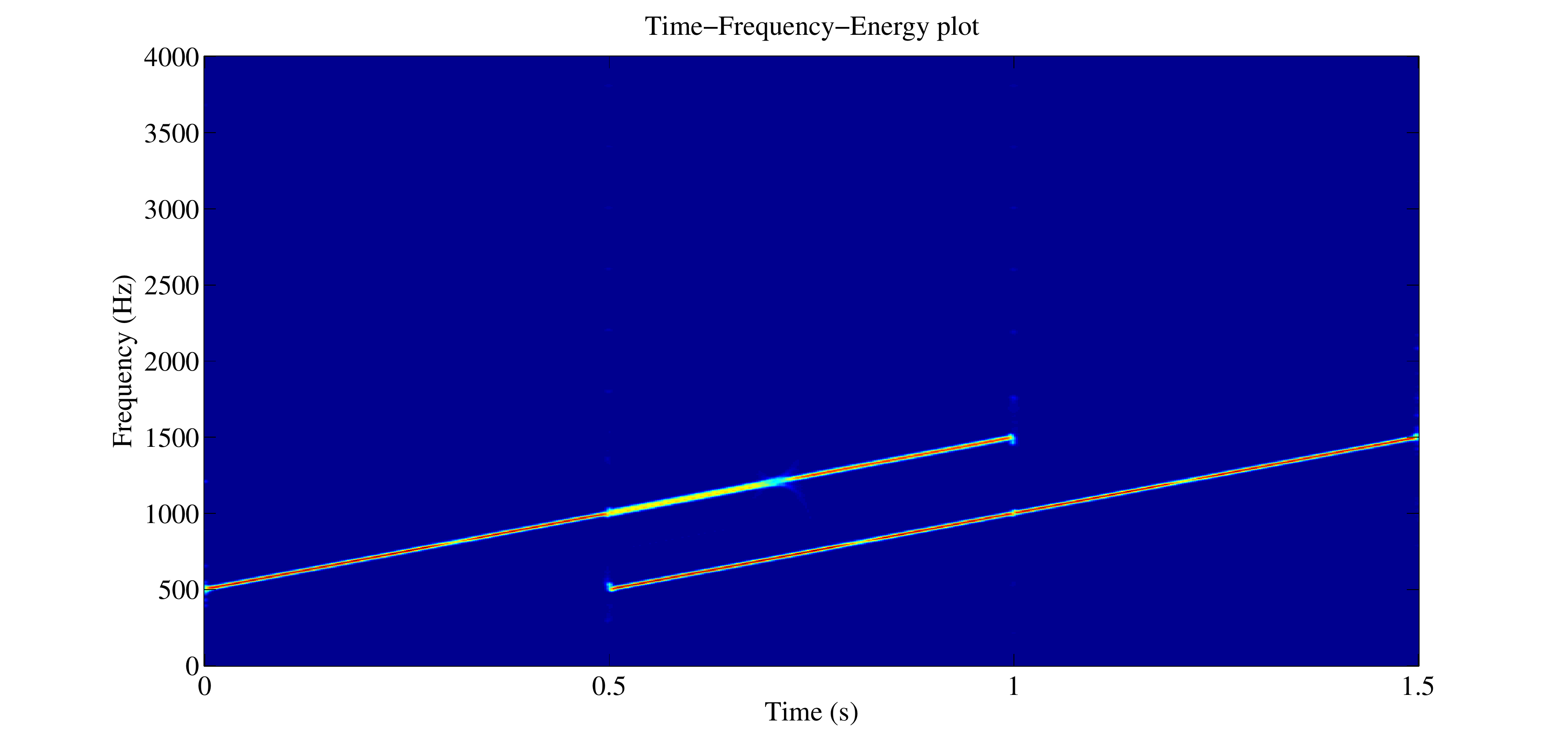}
\includegraphics[angle=0,width=0.5\textwidth,height=0.3\textwidth]{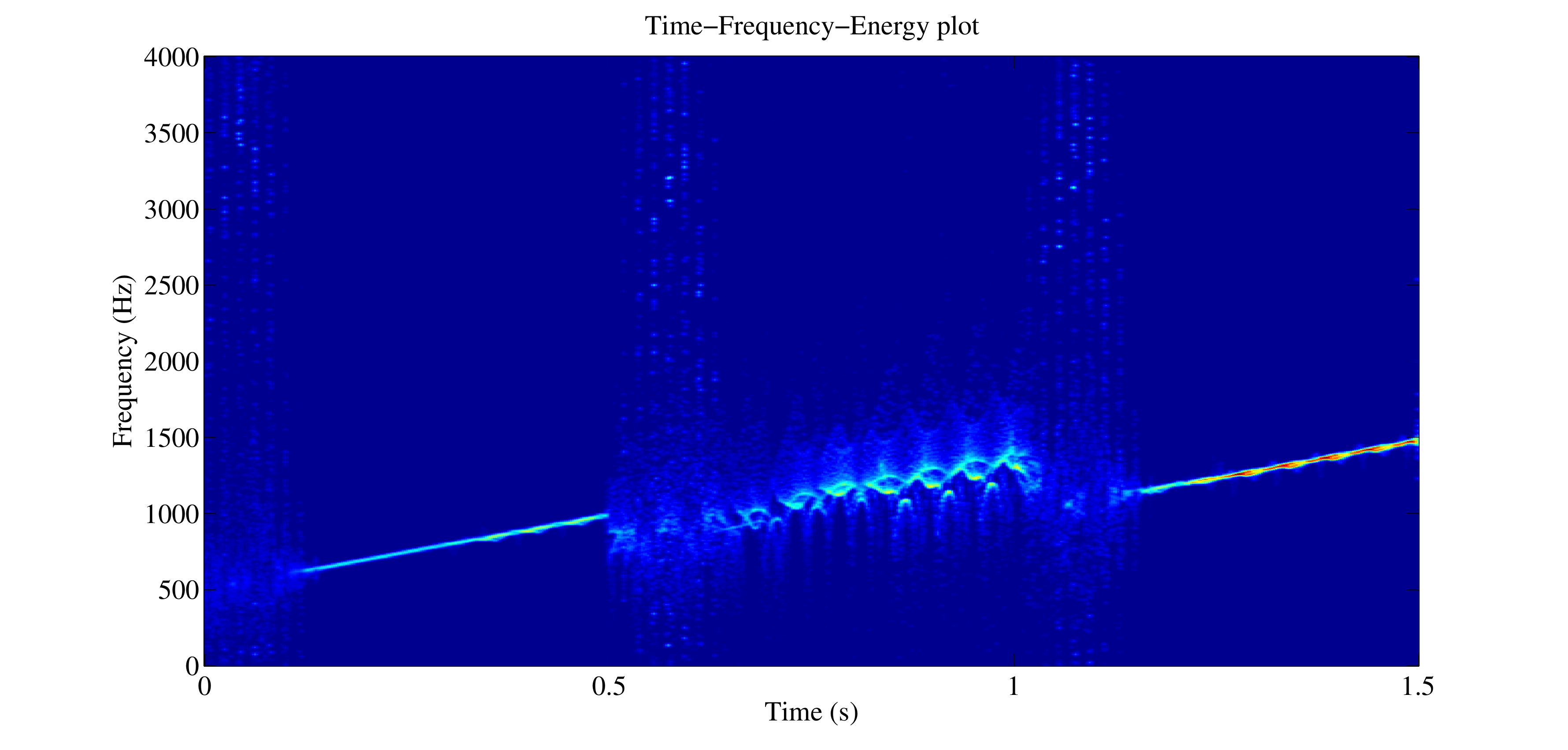}
\caption{The TFE analysis of a nonstationary signal, which is sum of a linear chirp signal and its delayed version, by this proposed method: (top figure) with zero-phase FIR filter based decomposition into 10 bands, and (bottom figure) with conventional FIR filter based decomposition into 10 bands of equal bandwidth.}
\label{fig:ChirpAndItsDelayed1}
\end{figure}

Figure~\ref{fig:ChirpAndItsDelayed1} shows the TFE distribution of a nonstationary signal (sum of a linear chirp signal and its delayed version) by this proposed IF: (top figure) with zero-phase FIR filter based decomposition into 10 bands; and (bottom figure) with conventional FIR filter based decomposition into 10 bands of equal bandwidth, which clearly not able to reveal true frequencies present in the signal.

\textbf{Discussion:} From Examples (1, 2 and 3), it is clear that the zero-phase DFT or zero-phase FIR filter based decomposition is able to track TFE distribution present in a signal. However, conventional filtering cannot be used to obtain meaningful TFE distribution of a signal, which is clearly demonstrated in bottom figures of Figure~\ref{fig:lcPlusFmp2fmd}, Figure~\ref{fig:ParallelChirps1FIR} and Figure~\ref{fig:ChirpAndItsDelayed1}.

\begin{figure}[!t]
\centering
\includegraphics[angle=0,width=0.5\textwidth,height=0.3\textwidth]{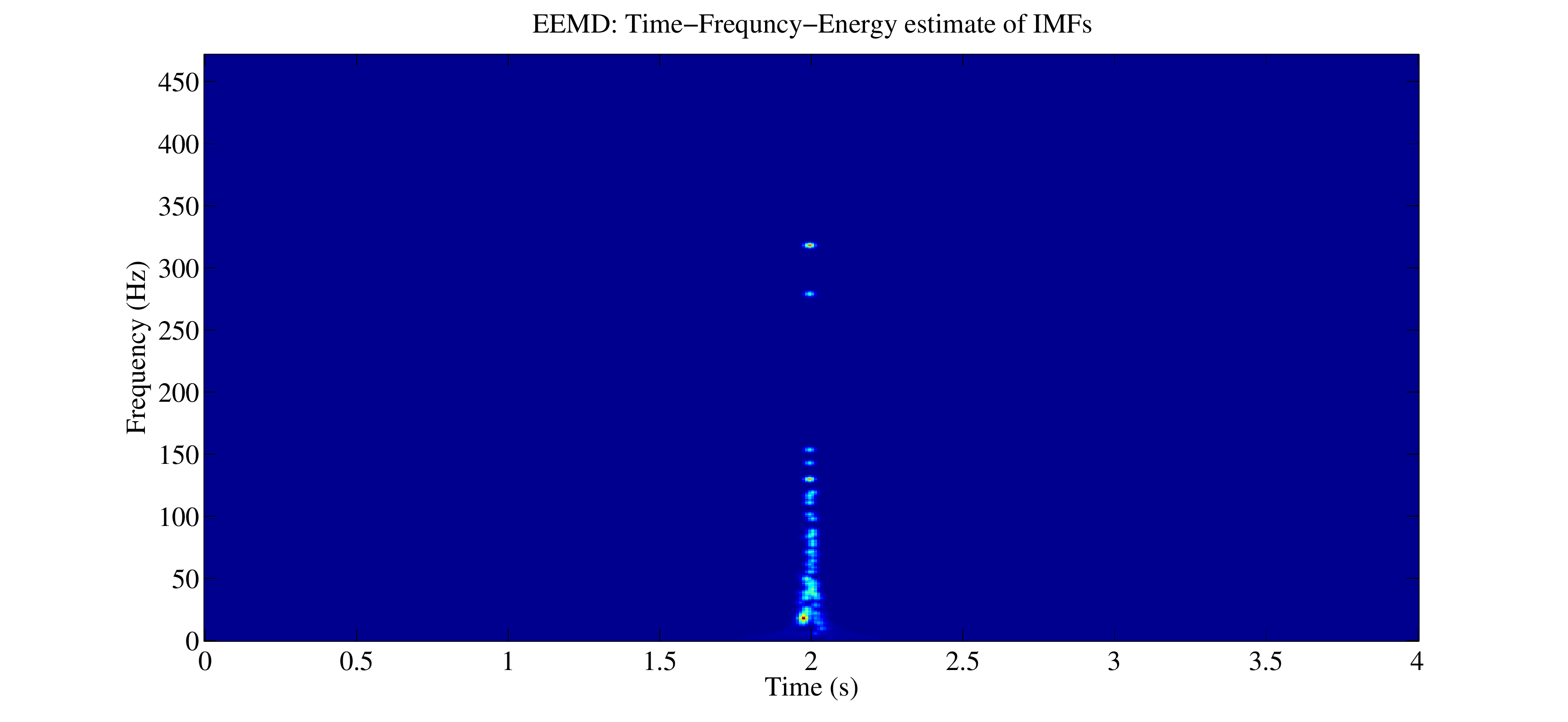}
\includegraphics[angle=0,width=0.5\textwidth,height=0.3\textwidth]{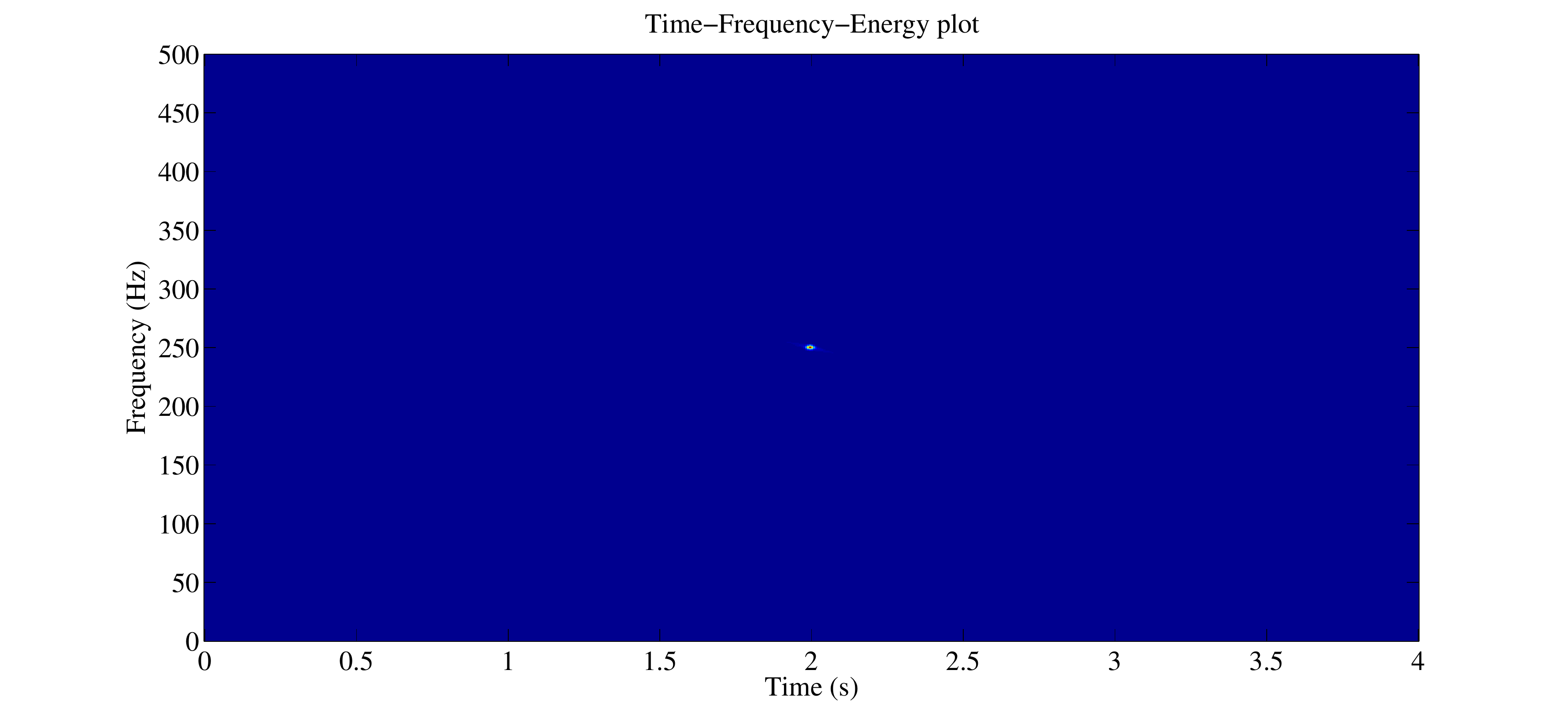}
\caption{The TFE analysis of unit sample sequence $\delta[n-n_0]$ (with, $n_0 = 1999$,
sampling frequency Fs = 1000 Hz, length $N = 4000$) by ensemble EMD (EEMD) (top figure) and this proposed method without decomposition (bottom figure).}
\label{fig:ui}
\end{figure}

\textbf{Example 4:} The unit sample sequence (delta function) is defined as $x[n]=\delta [n-n_0]=1$ at $n=n_0$ and zero otherwise. Using the discrete-time Fourier transform (DTFT) $X(\omega)=\sum_{n=-\infty}^{\infty} x[n] \exp(-j\omega n)$, one can obtain the DTFT of unit sample sequence as $X(\omega)=\exp(-j\omega n_0) \Rightarrow |X(\omega)|=1$. Using the inverse DTFT (IDTFT), $x[n]=\frac{1}{2\pi}\int_{-\pi}^{\pi} X(\omega) \exp(j\omega n) \ud \omega$, one can represent the unit sample sequence as $x[n]=\frac{1}{2\pi}\int_{-\pi}^{\pi} \exp(j\omega (n-n_0)) \ud \omega$.
This representation demonstrate that it is a superposition of equal amplitude sinusoidal functions of all frequencies [0--$\pi$). The Nyquist frequency ($F_s/2$) is the highest frequency that can be present at a given sampling rate, $F_s$, in a discrete-time signal.
The analytic representation of this signal is given by~\cite{rslc9} $z[n] =\frac{\sin(\pi (n-n_0))+j[1-\cos(\pi (n-n_0))]}{\pi (n-n_0)}=a[n]\exp(j\phi [n])$, where $a[n] =\left\lvert\frac{ \sin(\frac{\pi}{2} (n-n_0))} {\frac{\pi}{2} (n-n_0))}\right\rvert$, $\phi[n] =\frac{\pi}{2} (n-n_0)$, and real part of $z[n]$ is the original signal as $\frac{\sin(\pi (n-n_0))}{\pi (n-n_0)}=\delta [n-n_0]$. Thus, frequency $\omega[n] =\phi[n+1]-\phi[n]=\frac{\pi}{2}$, which corresponds to half of the Nyquist frequency, i.e. $F_s/4$.
Figure~\ref{fig:ui} shows the TFE estimates of this signal (with $n_0 = 1999$, $Fs = 1000$ Hz and length $N = 4000$) using the EMD (top figure) and proposed method without decomposition (bottom figure).
We again observe that the frequency present in this TFE plot is a average frequency (because delta function contains equal amplitude sinusoids of all frequencies 0 to $F_s/2$). This example also demonstrate that the TFE plot, obtained by the Hilbert spectrum, is not limited by uncertainty principle and signal can be highly concentrated in time and frequency plane. However, it is to be noted that this TFE plot is not providing the true frequencies of delta function.

\textbf{Discussion:} In order to explain a average frequency effect in Figure~\ref{fig:lcPlusFmp}, Figure~\ref{fig:ParallelChirps}, Figure~\ref{fig:ChirpAndItsDelayed} and Figure~\ref{fig:ui}, let us consider a sum of sinusoids of equal amplitudes  $x(t)=\sum_{k=1}^N A\cos(\omega_0 kt)$. Its analytic representation is given by $z(t)=\sum_{k=1}^N A\exp(j\omega_0 kt)=\frac{A\sin(\omega_0 \frac{N}{2}t)}{\sin(\omega_0 t/2)}\exp(j\omega_0\frac{N+1}{2}t)=a(t)\exp(j\phi (t))$ which implies phase $\phi (t)=(\omega_0\frac{N+1}{2}t)$ and hence IF $f(t)=\frac{1}{2\pi}\omega_0\frac{N+1}{2}$. This is what we observe in these figures (especially Figure~\ref{fig:ui} bottom one). Here, it is to be noted that if amplitudes of constituent sinusoids are not equal, then resultant IF would not be a constant (average) frequency but it would be a variable one.

\begin{figure}[!t]
\centering
\includegraphics[angle=0,width=0.5\textwidth,height=0.3\textwidth]{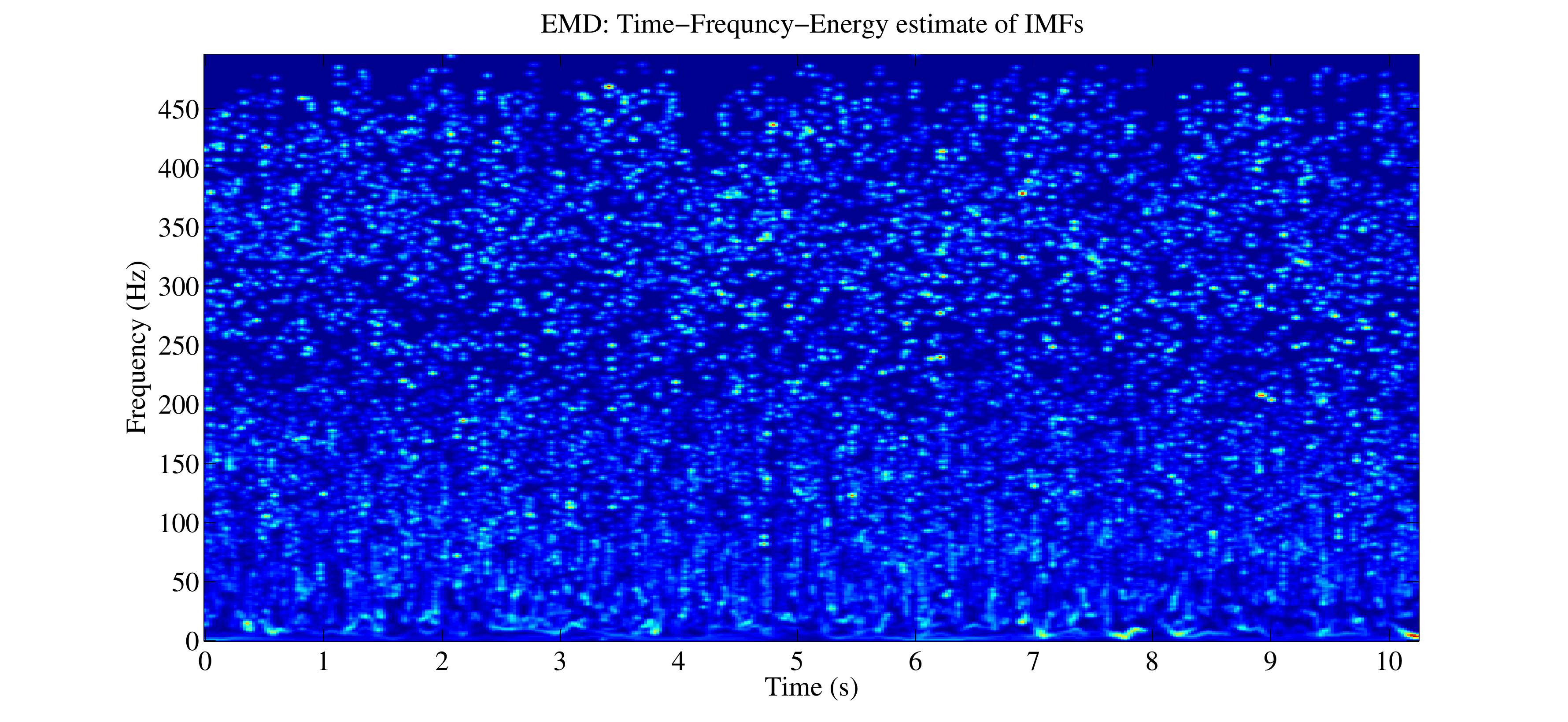}
\includegraphics[angle=0,width=0.5\textwidth,height=0.3\textwidth]{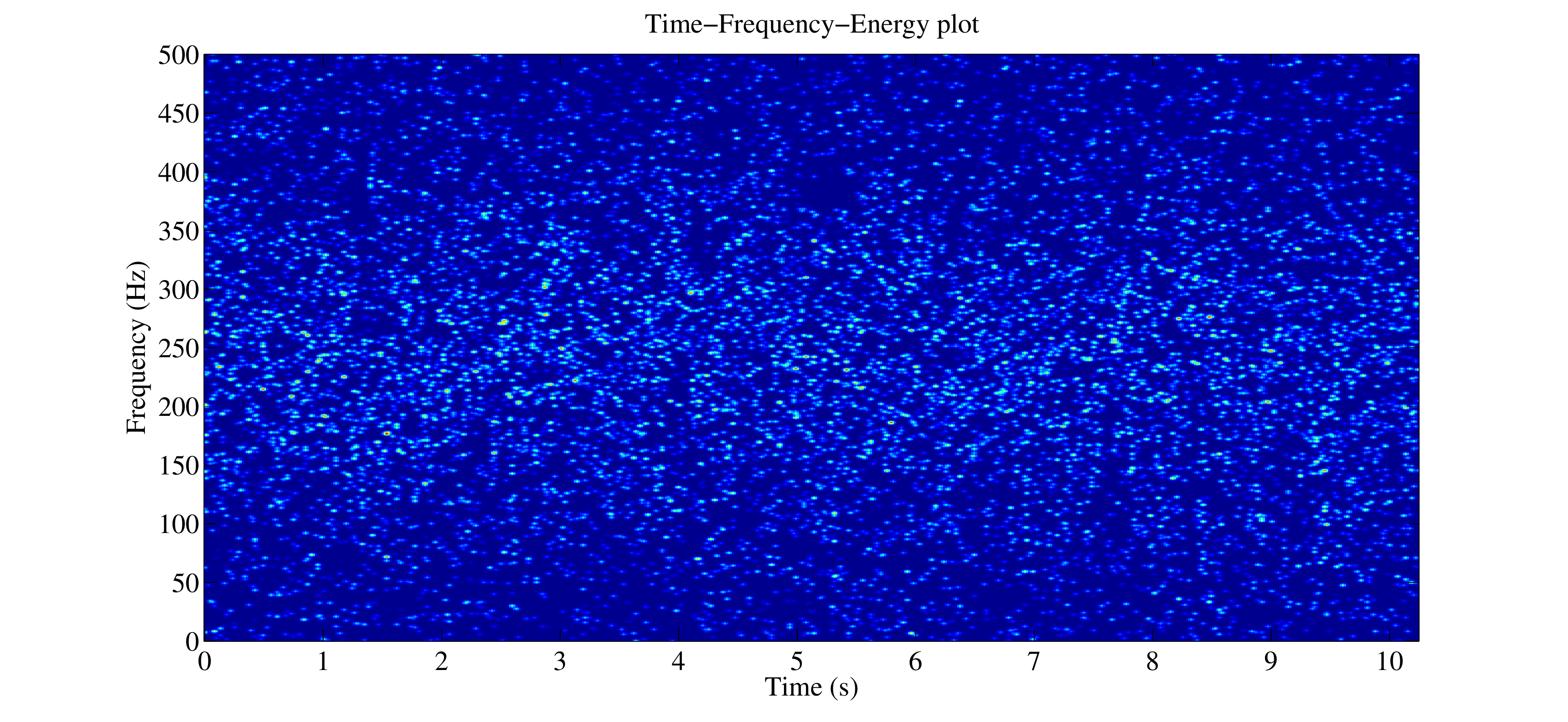}
\caption{The TFE analysis of the Gaussian white noise with zero mean and unit variance (with sampling frequency Fs = 1000 Hz, length $N = 10240$) by the EMD (top figure) and this proposed method without decomposition (bottom figure).}
\label{fig:randn}
\end{figure}

\textbf{Example 5:} Figure~\ref{fig:randn} shows the TFE analysis of a white Gaussian noise (with zero mean, unit variance, 10240 samples and sampling frequency $F_s=100$ Hz) obtained from the EMD (top one) and by the proposed method without decomposition (bottom one). It is clear that the energy in TFE plane is more concentrated at mid frequencies and almost randomly distributed across all the other frequency ranges by the proposed method. However, distribution of the energy in TFE plane is less random and concentration is weak at high frequencies by the EMD algorithm.

\begin{figure}[!t]
\centering
\includegraphics[angle=0,width=0.5\textwidth,height=0.3\textwidth]{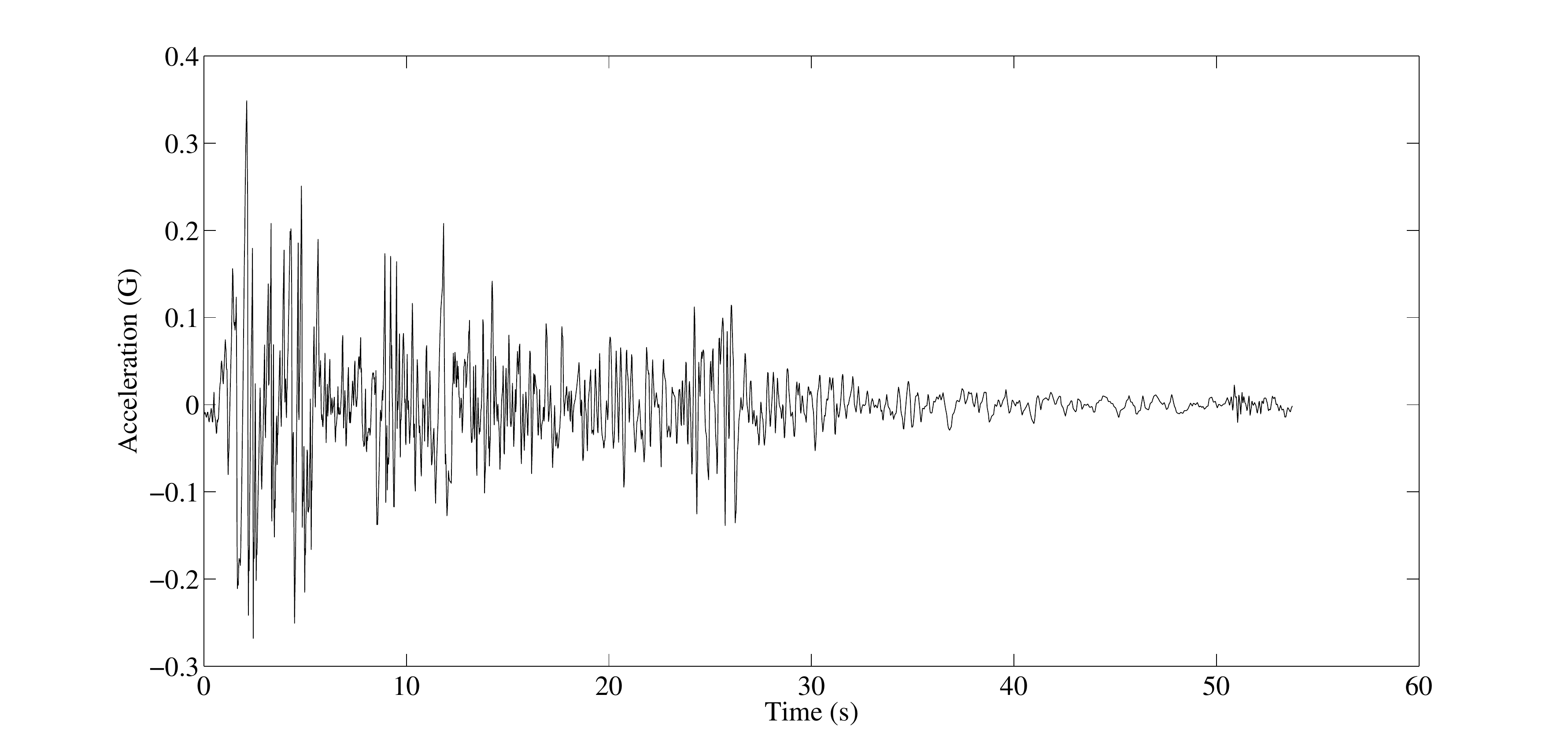}
\includegraphics[angle=0,width=0.5\textwidth,height=0.3\textwidth]{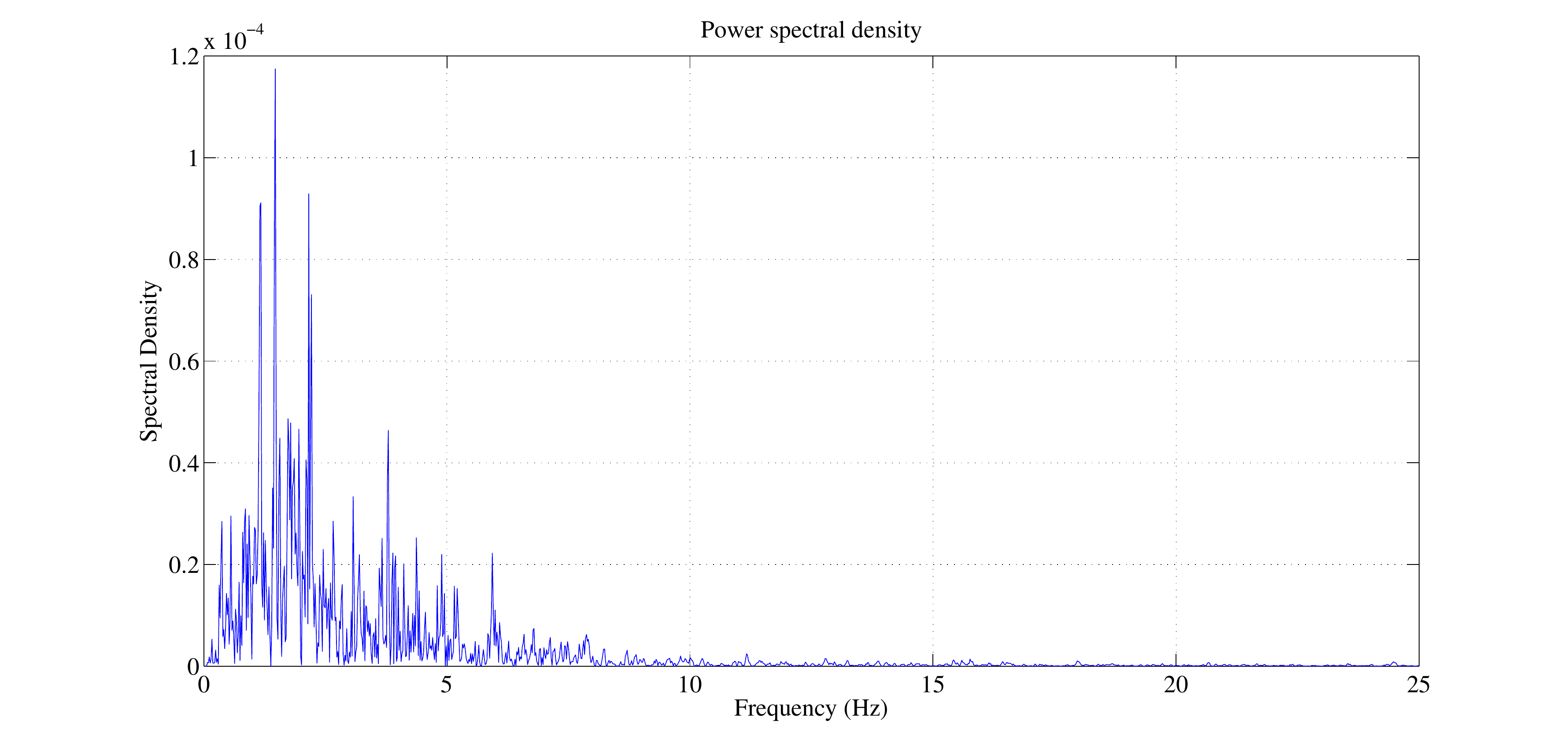}
\caption{The Elcentro Earthquake May 18, 1940 North-South Component data (top figure), Fourier based power spectral density (PSD) (bottom figure).}
\label{fig:eqtfe}
\end{figure}

\begin{figure}[!t]
\centering
\includegraphics[angle=0,width=0.5\textwidth,height=0.3\textwidth]{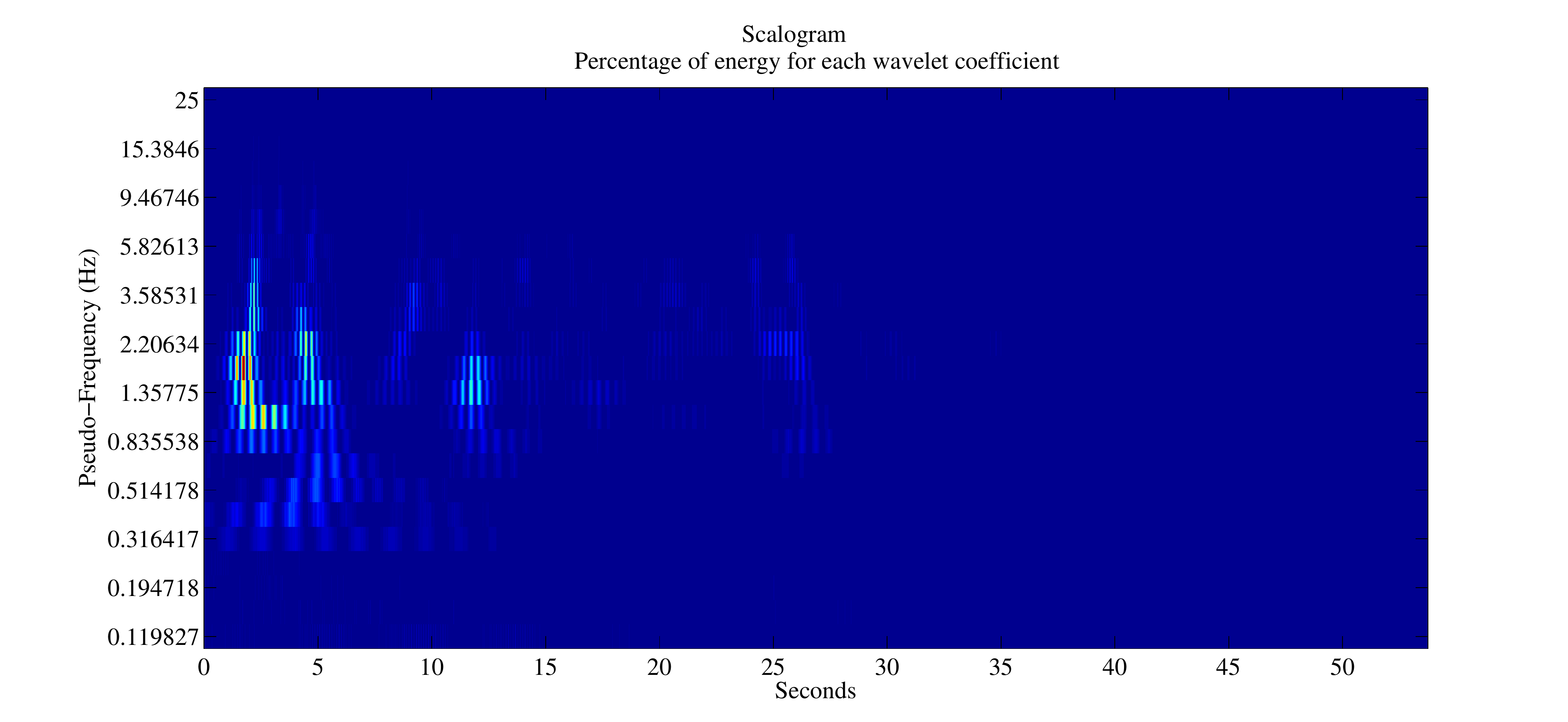}
\includegraphics[angle=0,width=0.5\textwidth,height=0.3\textwidth]{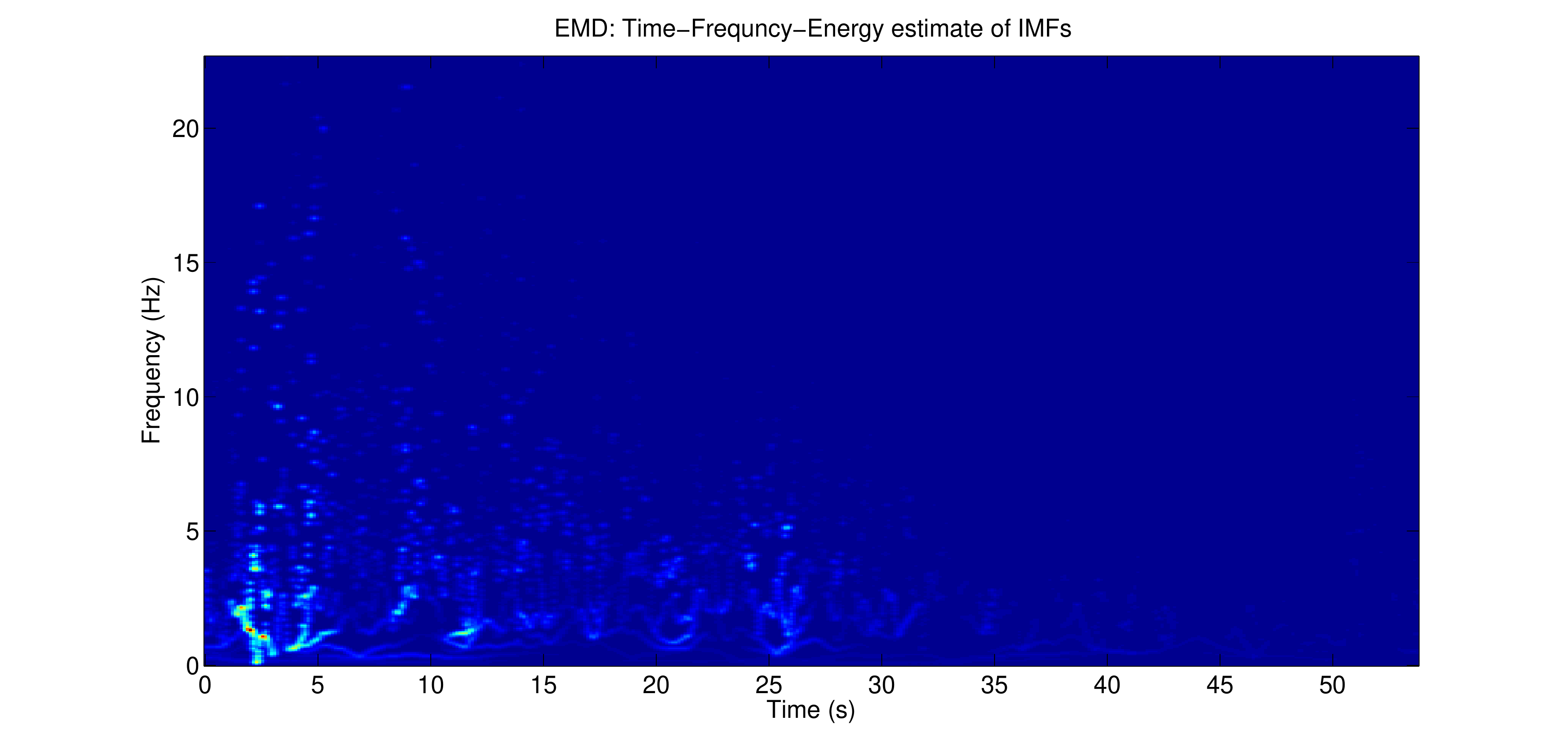}
\includegraphics[angle=0,width=0.5\textwidth,height=0.3\textwidth]{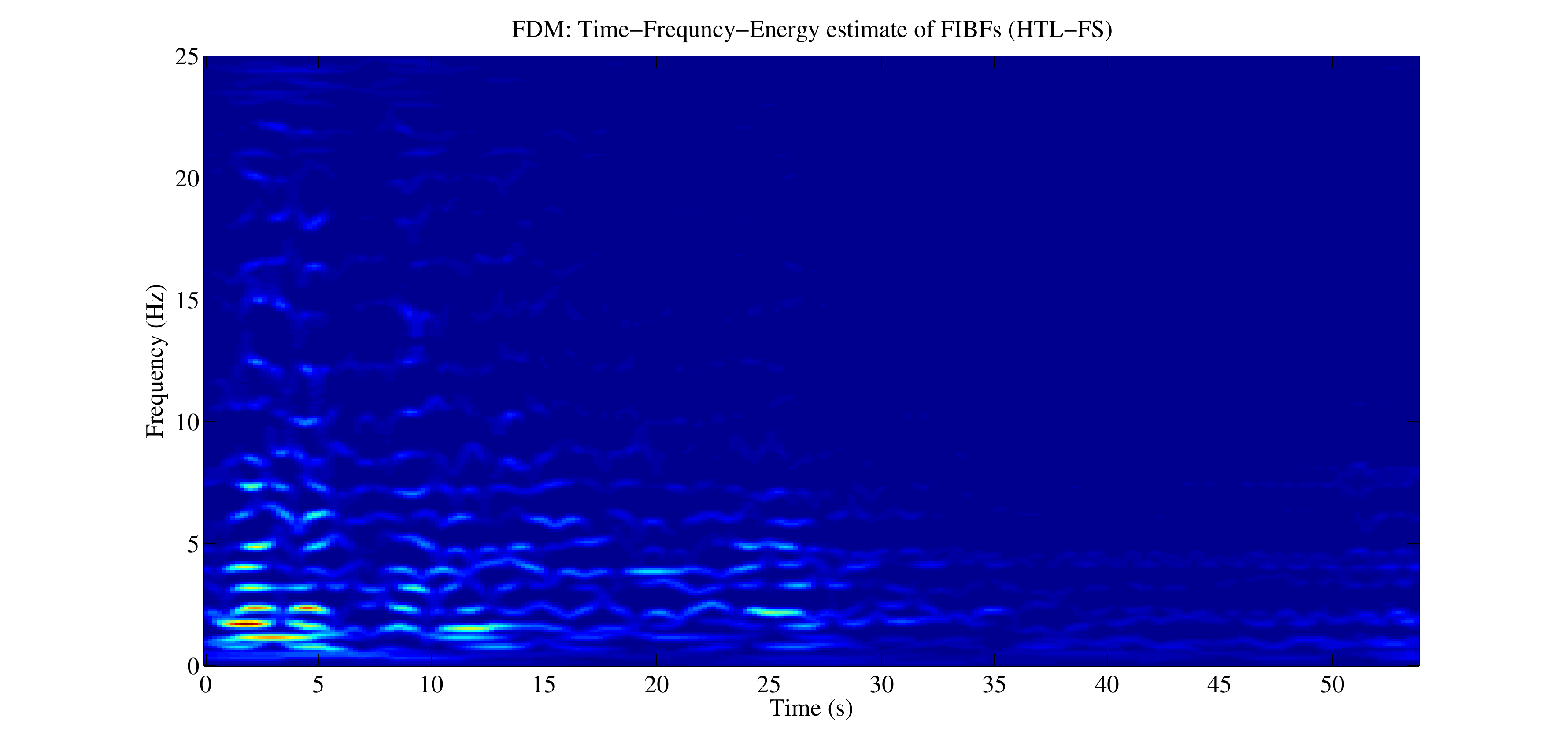}
\caption{The TFE plot of the Elcentro Earthquake data (top to bottom) using the: (a) CWT (b) EMD and (c) FDM.}
\label{fig:eq_TFE1}
\end{figure}

\begin{figure}[!t]
\centering
\includegraphics[angle=0,width=0.5\textwidth,height=0.3\textwidth]{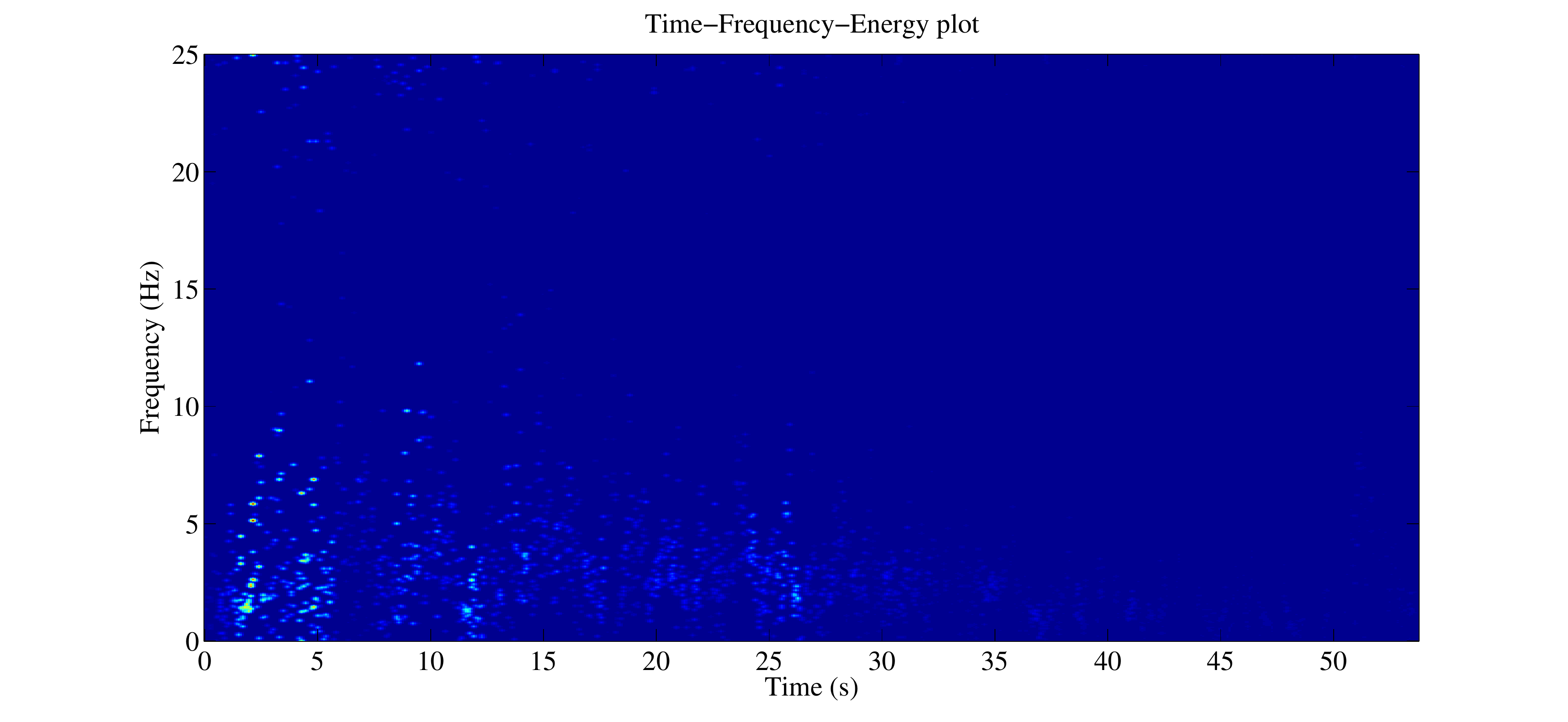}
\includegraphics[angle=0,width=0.5\textwidth,height=0.3\textwidth]{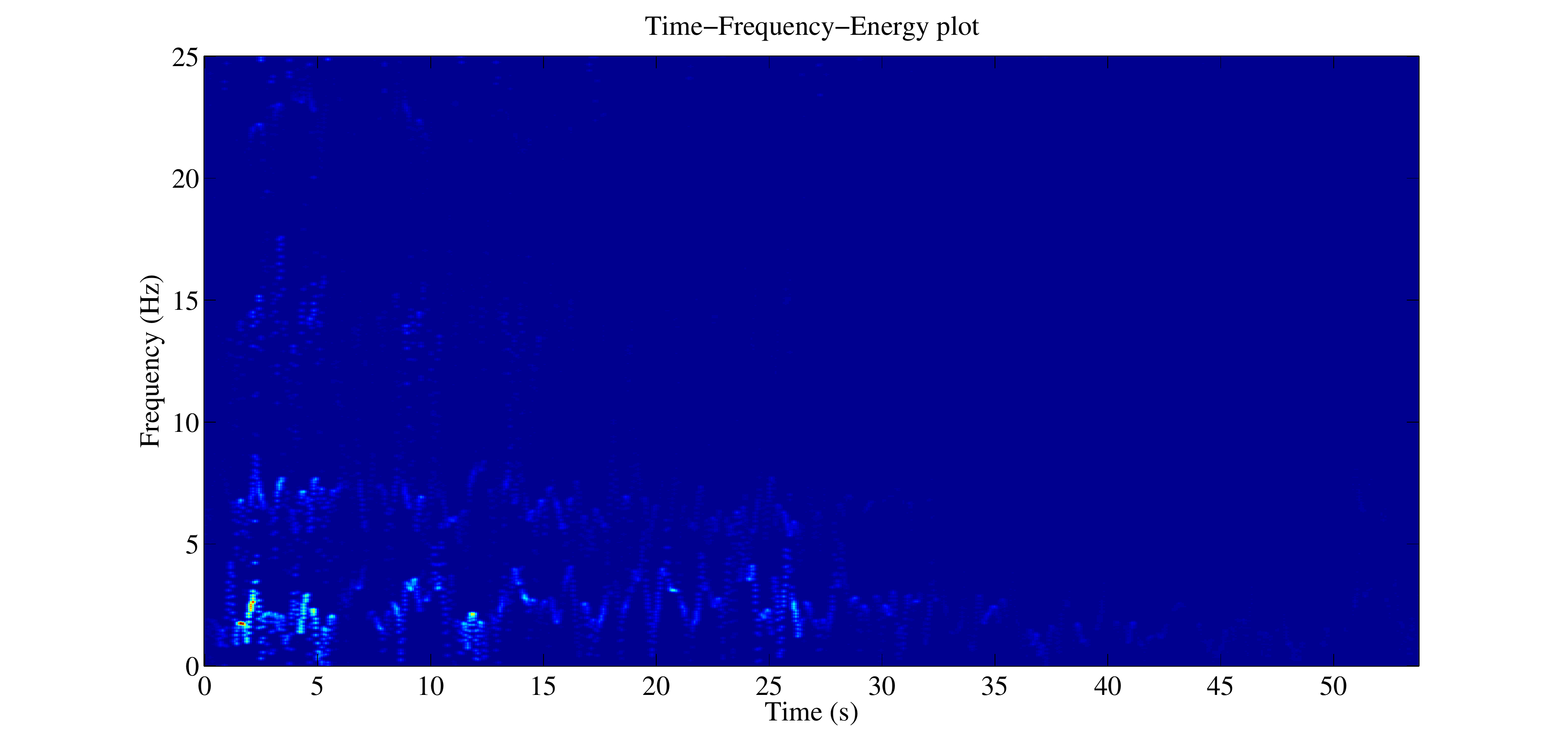}
\includegraphics[angle=0,width=0.5\textwidth,height=0.3\textwidth]{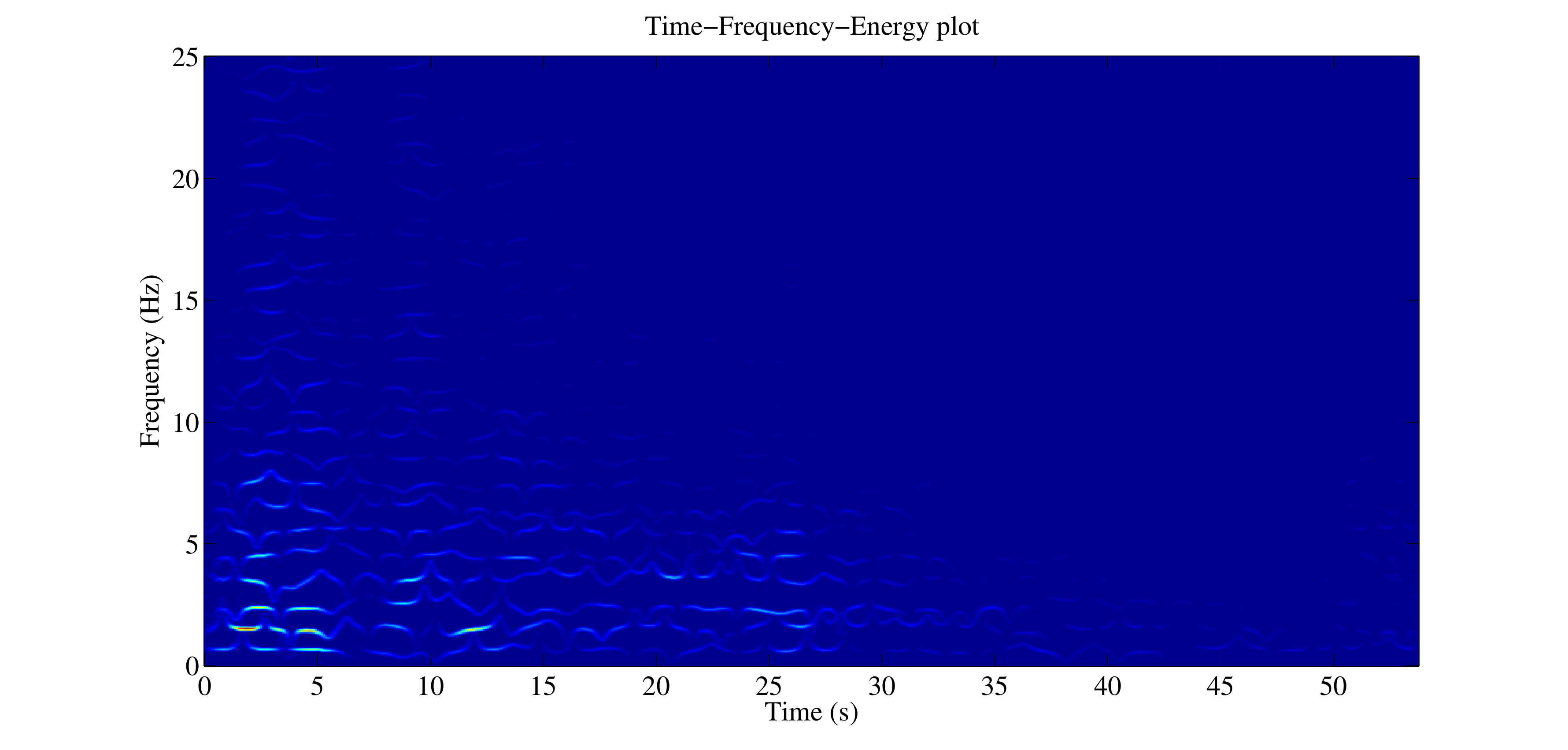}
\caption{The TFE plot of the Elcentro Earthquake data (top to bottom) using the: (a) proposed method without decomposition and (b) proposed method with DFT based decomposition into four bands [0--5, 5--10, 10--20, 20--25] Hz and (c) DFT based decomposition into 25 bands of 1 Hz each.}
\label{fig:eq_TFE2}
\end{figure}

\textbf{Example 6:} An Earthquake time series is a nonlinear and nonstationary data. The Elcentro Earthquake data (sampled at $F_s= 50Hz$) has been taken from~\cite{EQ33} and is shown in Figure~\ref{fig:eqtfe} (top one). The critical frequency range that matter in the structural design is less than $10Hz$, and the Fourier based power spectral density (PSD), Figure~\ref{fig:eqtfe} (bottom one), show that almost all the energy in this data is within $10Hz$.
The TFE distributions by the (a) continuous wavelet transform (CWT) (b) EMD and (c) FDM methods are shown in Figure~\ref{fig:eq_TFE1}.
The TFE distributions by the proposed method (a) without decomposition (b) with DFT based decomposition into four bands [0--5, 5--10, 10--20, 20--25] Hz and (c) with DFT based decomposition into 25 bands of 1 Hz are shown in Figure~\ref{fig:eq_TFE2}. These TFE distribution indicate that the maximum energy concentration is around $1.7Hz$ and 2 second. The TFE plot provide details of how the different waves arrive from the epical center to the recording station, e.g. the compression waves of small amplitude but higher frequency range of $10$ to $20Hz$, the shear and surface waves of strongest amplitude and lower frequency range of below $5Hz$ which does most of the damage, and other body shear waves which are present over the full duration of the data span.

\begin{figure}[!t]
\centering
\includegraphics[angle=0,width=0.5\textwidth,height=0.3\textwidth]{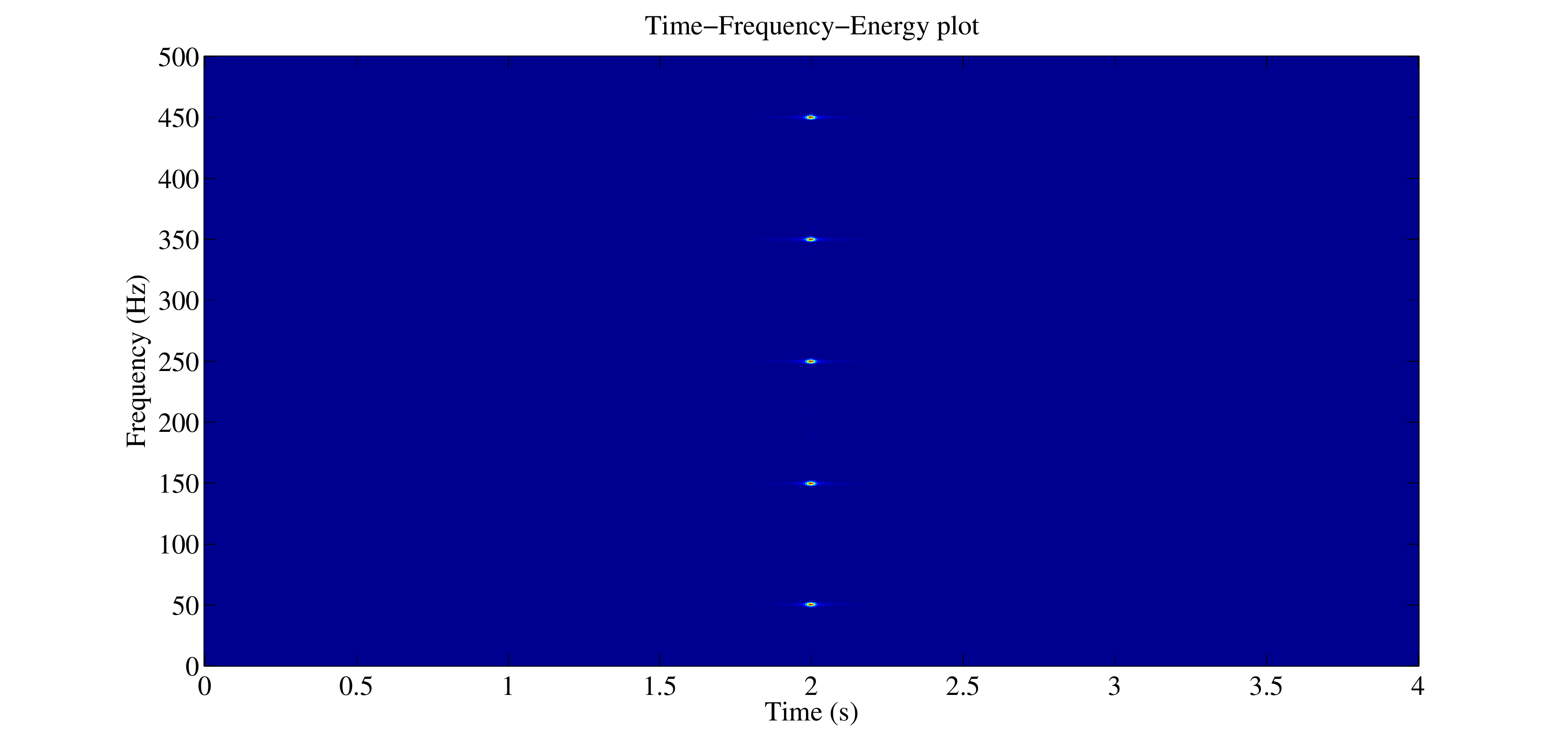}
\includegraphics[angle=0,width=0.5\textwidth,height=0.3\textwidth]{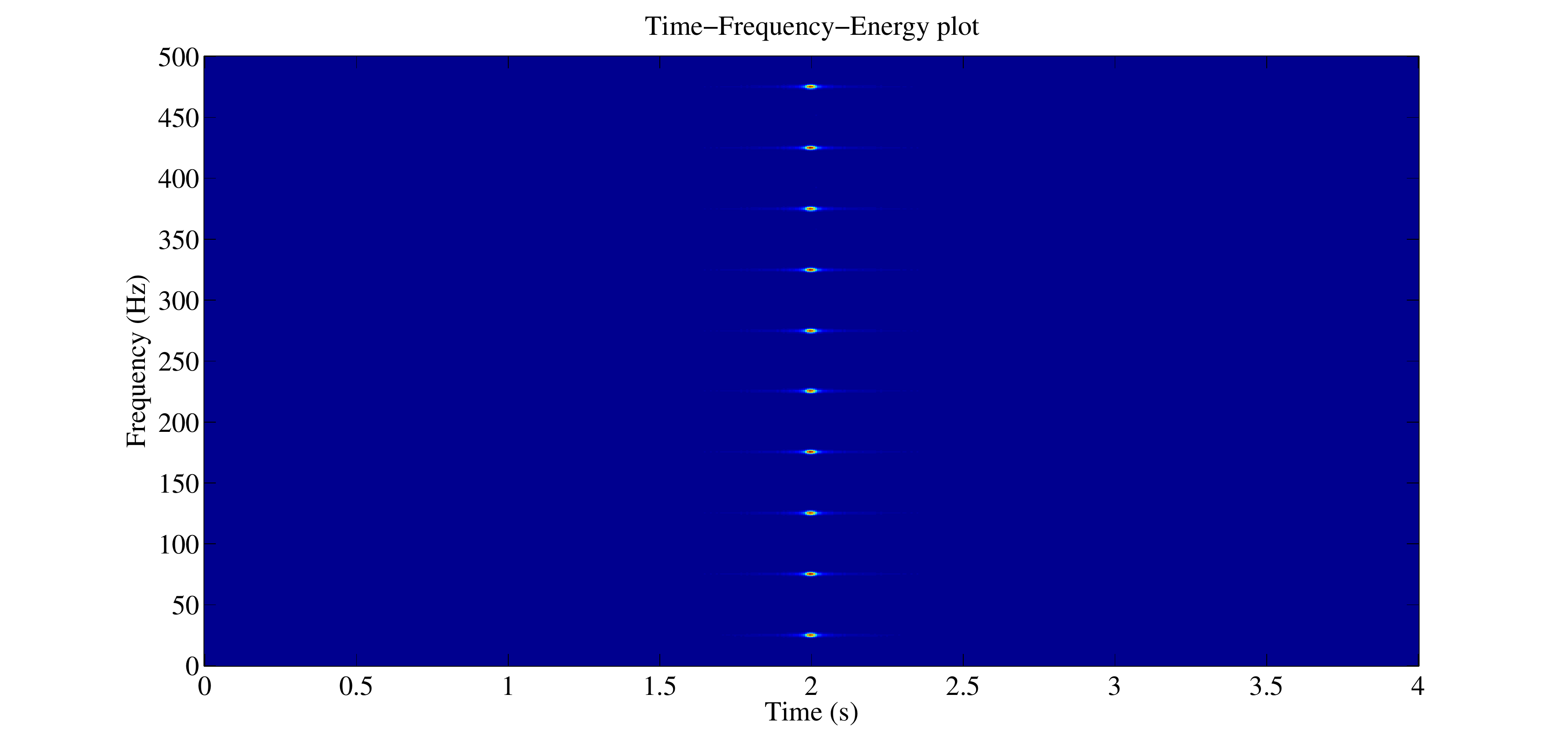}
\caption{The TFE analysis of unit sample sequence $\delta[n-n_0]$ (with, $n_0 = 1999$,
sampling frequency Fs = 1000 Hz, length $N = 4000$) by proposed method with DFT based decomposition into five (top figure) and ten (bottom figure) bands of equal bandwidth.}
\label{fig:ui105}
\end{figure}

\textbf{Discussion:} It is clear from the above examples that different methods are producing the different TFE distributions of a signal. So, before concluding anything one needs to be careful while doing the analysis. For example, frequencies present in the Fourier spectrum is telling that these frequencies are present all the time in a signal under analysis, which may not be true. Consider another example, Figure~\ref{fig:ui} (bottom), of unit sample sequence which is telling that this signal is concentrated in time (which is true) and also concentrated in frequency (which is not true). If this were true, then we could transmit delta function in almost zero time with zero bandwidth through any system or channel. Thus, we conclude that the TFE representation depends on the number of bands in which data has been divided. In order to further illustrate this point we divided the same data, which is used for Figure~\ref{fig:ui}, by DFT based zero-phase filtering into five and ten bands to obtain top and bottom figures in Figure~\ref{fig:ui105}, respectively, which reveal many those frequencies of data which are not present in Figure~\ref{fig:ui}.

From the Figures~(\ref{fig:lcPlusFmp},~\ref{fig:ParallelChirps},~\ref{fig:ChirpAndItsDelayed},~\ref{fig:ui} and~\ref{fig:ui105}), it is clear that the proposed method using the Hilbert spectrum produces average frequencies and good time resolution when the envelope of signal is smooth. However, if the envelope of a signal is fluctuating randomly or rapidly, e.g. consider the case of Gaussian white noise and Earthquake time series, the TFE plot has good time and frequency resolution as shown in Figure~\ref{fig:randn} and Figure~\ref{fig:eq_TFE2} (top one). It is also clear, form Figures~(\ref{fig:lcPlusFmp1},~\ref{fig:ParallelChirps},~\ref{fig:ChirpAndItsDelayed},~\ref{fig:ui} and~\ref{fig:ui105}), that as the signal is decomposed into more number of narrow bands, true frequencies present in the signal under analysis are revealed, frequency resolution is also increasing while the time resolution is reducing marginally.

\section{CONCLUSION}\label{con}
The instantaneous frequency (IF) is an important parameter for the analysis of nonstationary signals and nonlinear systems. It is the basis of the time-frequency-energy (TFE) analysis of a signal. The IF is the time derivative of the instantaneous phase and, originally, it is well-defined only when this derivative is positive. That is, the IF is valid only for monocomponent signals. If time derivative of instantaneous phase is negative, i.e. the IF is negative, then it does not provide any physical significance. This study proposed a mathematical solution and eliminate this problem by modifying the present definition of IF. This is achieved by using the property of the \emph{multivalued} inverse tangent function that provides base to ensure that the instantaneous phase is an increasing (or a nondecreasing) function.

There are two fundamental and important conceptual innovations of this work. First, the extension of the conventional definition of IF by redefining it such that it is always positive. This proposed IF is valid for all types of signals such as monocomponent and multicomponent, narrowband and wideband, stationary and nonstationary, linear and nonlinear signals. The understanding of the time-frequency-energy representation, by all the methods which are using the IF, would improve significantly by using this definition. Second, we have also demonstrated that the zero-phase filtering based decomposition of a signal into a set of desired frequency bands with proposed IF accurately reveals the TFE distribution. Whereas, conventional (non zero-phase) filtering based decomposition cannot be used to obtain correct and meaningful TFE distribution. The Fourier and filter theory are well established, fully matured and developed, thus the zero-phase filtering based decomposition of a signal is most powerful which presents full control over the number of bands with desired cutoff frequencies. This kind of control and features are difficult to achieve or may not be possible by the decomposition methods such as empirical mode decomposition (EMD) algorithms, synchrosqueezed wavelet transforms (SSWT), variational mode decomposition (VMD), eigenvalue decomposition (EVD), time-varying vibration decomposition, resonance-based signal decomposition, EMD based on constrained optimization and empirical wavelet transform (EWT) available in the literature.
Simulations and numerical results demonstrated the superiority, validity and efficacy of the proposed IF for the TFE analysis of a signal as compared to other existing methods available in the literature.
\section*{ACKNOWLEDGMENTS}
Author would like to show his gratitude to the Prof. SD Joshi (IITD), Prof. RK Pateny (IITD) and Dr. Kaushik Saha (Director, Samsung R\&D Institute India--Delhi) for sharing their wisdom and expertise during the course of this research.


\begin{thebibliography}{40}
\itemsep0em % to remove space between references
\bibitem{rslc1} {Huang N. E., Shen Z., Long S., Wu M., Shih H., Zheng Q., Yen N., Tung C., and Liu H.}, {The empirical mode decomposition and Hilbert spectrum for non-linear and non-stationary time series analysis}, \emph{Proc. R. Soc. A}, 454, 903--995, 1988.
\bibitem{blBB} Boashash B., Time Frequency Signal Analysis and Processing: A Comprehensive Reference, Elsevier, Boston, 2003.
\bibitem{rslc2} {Wu Z. and Huang N. E.}, {Ensemble Empirical Mode Decomposition: a noise-assisted data analysis method}, \emph{Adv. Adapt. Data Anal.}, 1 (1), 1--41, 2009.
\bibitem{rslc3} {Rehman N. and Mandic D. P.}, {Multivariate empirical mode decomposition}, \emph{Proc. R. Soc. A}, 466, 1291--1302, 2010.
\bibitem{rslc4} {Singh P., Joshi S. D., Patney R. K., Saha K.}, {The Hilbert spectrum and the Energy Preserving Empirical Mode Decomposition}, \emph{	 arXiv:1504.04104 [cs.IT]}, 2015.
\bibitem{rslc5} {Singh P., Joshi S. D., Patney R. K., Saha K.}, {Some studies on nonpolynomial interpolation and error analysis}, \emph{Applied Mathematics and Computation}, 244, 809--821, 2014.
\bibitem{rslc6} {Singh P., Srivastava P. K., Patney R. K., Joshi S. D., Saha K.}, {Nonpolynomial spline based empirical mode decomposition}, \emph{Signal Processing and Communication (ICSC), 2013 International Conference on}, 435--440, 2013.
\bibitem{rslc7} {Singh P., Joshi S. D., Patney R. K., Saha K.}, {The Linearly Independent Non Orthogonal yet Energy Preserving (LINOEP) vectors}, \emph{	 arXiv:1409.5710 [math.NA]}, 2014.

\bibitem{rslc71} Daubechies I., Lu J., Wu H. T., Synchrosqueezed Wavelet Transforms: an Empirical Mode Decomposition-like Tool, \emph{Appl. Comput. Harmon. Anal.}, 30, 243--261, 2011.
\bibitem{rslc72} Dragomiretskiy K., Zosso D., Variational Mode Decomposition, \emph{IEEE Transactions on Signal Processing}, 62 (3), 531--544, 2014.
\bibitem{rslc73} P. Jain, R. B. Pachori, An iterative approach for decomposition of multi-componentnon-stationary signals based on eigenvalue decomposition of the Hankel matrix, \emph{Journal of the Franklin Institute}, 352 (10), 4017--4044, 2015.
\bibitem{rslc74} Gilles J., Empirical Wavelet Transform, \emph{IEEE Transactions on Signal Processing},  61 (16), 3999--4010, 2013.

\bibitem{rslc8} Singh P., {Some studies on a generalized fourier expansion for nonlinear and nonstationary time series analysis}, \emph{PhD thesis}, department of electrical enigineering, IIT Delhi, India, 2016.
\bibitem{co1} Meignen S. and Perrier V., A new formulation for empirical mode decomposition based on constrained optimization, \emph{IEEE Signal Process. Lett.}, 14 (12), 932--935, 2007.
\bibitem{co2} Hou T. Y. and Shi Z., Adaptive data analysis via sparse time-frequency representation, \emph{Adv. Adapt. Data Anal.},  3 (1 \& 2), 1--28, 2011.
\bibitem{co3} Feldman M., Time-varying vibration decomposition and analysis based on the Hilbert transform, \emph{J. Sound Vibrat.}, 295(3--5), 518--530, 2006.
\bibitem{co4} Selesnick I. W., Resonance-based signal decomposition: A new sparsity-enabled signal analysis method, \emph{Signal Process.}, 91 (12), 2793--2809, 2011.
\bibitem{rslc9} {Singh P., Joshi S. D., Patney R. K., Saha K.}, {The Fourier Decomposition Method for nonlinear and non-stationary time series analysis}, \emph{arXiv:1503.06675 [stat.ME]}, 2015.
%\bibitem{rslc10} Singh P., Joshi S. D., Patney R. K., Saha K., {The Taylor's nonpolynomial series approximation}, 2016, hal-01229594v2.
\bibitem{rslc101} Singh P., {Time-Frequency analysis via the Fourier Representation}, \emph{arXiv:1604.04992 [cs.IT]}, 2016.
\bibitem{th46} {Cummings D. A., Irizarry R. A., Huang N.E., Endy T. P., Nisalak A., Ungchusak K., Burke D. S.}, {Travelling waves in the occurrence of dengue haemorrhagic fever in Thailand}, \emph{Nature}, 427, 344--347, 2004.
\bibitem{rslc11} Singh P., Joshi S. D., {Some studies on multidimensional Fourier theory for Hilbert transform, analytic signal and space-time series analysis,} \emph{arXiv:1507.08117 [cs.IT]}, 2015.

\bibitem{rslc12}  Singh P., {LINOEP vectors, spiral of Theodorus, and nonlinear time-invariant system models of mode decomposition}, \emph{arXiv:1509.08667 [cs.IT]}, 2015.

\bibitem{Gabor} Gabor D., Theory of communication, Proc. IEE, 93 (III), 429--457, 1946.

\bibitem{rslc13} {Singh P., Joshi S. D., Patney R. K., Saha K.}, Fourier-based Feature Extraction for Classification of EEG Signals Using EEG Rhythms, \emph{Circuits, Systems, and Signal Processing}, DOI 10.1007/s00034-015-0225-z, 2015.
\bibitem{th19} Carson J., Fry T., Variable frequency electric circuit theory with application to the theory of frequency modulation, \emph{Bell System Tech. J.}, 16, 513--540, 1937.
\bibitem{th20} Van der Pol B., The fundamental principles of frequency modulation, \emph{Proc. IEE}, 93 (111), 153--158, 1946.
\bibitem{th21} Gabor D., Theory of communication, \emph{Proc. IEE}, 93 (III), 429--457, 1946.
\bibitem{th22} Hildebrand F. B., Advanced Calculus for Engineers, Englewood Cliffs, NJ: Prentice-Hall, 1949.
\bibitem{th23} Ville J., Theorie et application de la notion de signal analytic, Cables et Transmissions, 2A (1), 61-74, Paris, France,
1948. Translation by I. Selin, Theory and applications of the notion of complex signal, Report T-92, RAND Corporation, Santa Monica, CA.
\bibitem{th4} Boashash B., Estimating and interpreting the instantaneous frequency of a signal--Part 1: Fundamentals, \emph{Proc. IEEE}, 80 (4), 520--538, 1992.
\bibitem{th411} Boashash B., Estimating and interpreting the instantaneous frequency of a signal--Part 2: Algorithms and Applications, \emph{Proc. IEEE}, 80 (4), 540--568, 1992.
\bibitem{matlabweb} [Online:] \url{http://in.mathworks.com/help/matlab/ref/unwrap.html}.
\bibitem{EQ33} [Online:] \url{http://www.vibrationdata.com/elcentro.htm}.
\bibitem{rs22} {Mandic D. P., Rehman N., Wu Z. and Huang N.E.}, {Empirical Mode Decomposition-Based Time-Frequency Analysis of Multivariate Signals}, \emph{IEEE signal processing magazine}, (2013) November, 74--86.
 \bibitem{MTCode}[Online:] \url{https://www.researchgate.net/publication/307606777_MATLABCodeOfBreakingTheLimitsRedefiningTheIF}.
\end{thebibliography}
\end{document}